\documentclass[
 aip,
 amsmath,amssymb,
 reprint,
]{revtex4-1}

\usepackage{graphicx}
\usepackage{dcolumn}
\usepackage{bm}

\usepackage[utf8]{inputenc}
\usepackage[T1]{fontenc}
\usepackage{mathptmx}
\usepackage{etoolbox}
\usepackage{xcolor}

\makeatletter
\def\@email#1#2{
 \endgroup
 \patchcmd{\titleblock@produce}
  {\frontmatter@RRAPformat}
  {\frontmatter@RRAPformat{\produce@RRAP{*#1\href{mailto:#2}{#2}}}\frontmatter@RRAPformat}
  {}{}
}
\makeatother
\begin{document}

\preprint{AIP/123-QED}

\title[Journal of Rheology]{Time-resolved microstructural changes in large amplitude oscillatory shear of model single and double component soft gels}

\author{Gavin J. Donley}
 \altaffiliation[Also at ]{Infrastructure Materials Group, Materials and Structural Systems Division, Engineering Laboratory, National Institute of Standards and Technology, Gaithersburg, MD 20899, USA}
\author{Minaspi Bantawa}
 \altaffiliation[Now at ]{McKetta Department of Chemical Engineering, University of Texas at Austin, Austin, Texas 78712, USA}
\author{Emanuela Del Gado}
 \email{ed610@georgetown.edu}
\affiliation{Department of Physics \& Institute for Soft Matter Synthesis and Metrology, Georgetown University, Washington, DC 20057, USA}%

\date{\today}

\begin{abstract}
Soft particulate gels can reversibly yield when sufficient deformation is applied, and the characteristics of this transition can be enhanced or limited by designing hybrid hydrogel composites. While the microscopic dynamics and macroscopic rheology of these systems have been studied separately in detail, the development of direct connections between the two has been difficult, particularly with regard to the non-linear rheology. To bridge this gap, we perform a series of large amplitude oscillatory shear (LAOS) numerical measurements on model soft particulate gels at different volume fractions using coarse-grained molecular dynamics simulations. We first study a particulate network with local bending stiffness and then we combine it with a second component that can provide additional crosslinking to obtain two-component networks. 
Through the sequence of physical processes (SPP) framework we define time-resolved dynamic moduli and, by tracking the changes in these moduli through the period, we can distinguish transitions in the material behavior as a function of time. This approach helps us  establish the microsopic origin of the non-linear rheology by connecting the changes in dynamic moduli to the corresponding microstructural changes during the deformation including the non-affine displacement of particles, and the breakage, formation, and orientation of bonds.
\end{abstract}

\maketitle

\section{\label{intro} Introduction}

Soft particulate gels can result from a range of colloidal suspensions of particles, particle aggregates, fibrils or even droplets \cite{petekidis_wagner_2021, poon-pusey, trappe01, laurati09, helgeson14}. Attractive inter-particle interactions drive the aggregation  of structures that lead to kinetic arrest and result in the self-assembly of interconnected space-spanning networks at low particle volume fractions with solid-like elastic properties. These gels are used in a variety of industries, including foods, consumer products and biotechnologies. The highly tunable and adaptive mechanics and pronounced viscoelasticity of these systems make them useful in a number of applications including tissue regeneration scaffolds, drug delivery, electronics and battery technologies \cite{Hoffman2002,GUO2019}. However, the disorder and complex organization of gel constituents in a variety of mesoscopic structures can result in complex relaxation processes that pose a formidable challenge towards understanding the interplay between structure, dynamics and mechanics \cite{cipelletti2005slow, bouzid2017elastically, perge2014time, KeshavarzPNAS2021, aime2018microscopic}. For example, the rheological properties of the gel can be changed substantially as a result of the gelation pathway to the final structure, of the aging of the structure after gelation, and of the presence of multiple gel components. Hence understanding the changes in the rheological response, and their microstructural and kinetic origin, may help unlock new design principles. In addition, the incorporation of new constituents with different chemistry, surface properties, or gel interactions, may lead to distinct microscopic dynamics and mechanics and to the possibility to design composite gels with specific and new functionalities \cite{Haraguchi2002,Wu2009,Gu2017,vereroudakis20,multi-gels}. Designing materials with such controlled and tunable physical and mechanical properties can pave the way for obtaining smart materials with versatile features such as self-healing, thermo-responsiveness, toughness and extensibility \cite{Miao2015,Li2017,Gong2003,Filippidi2017,Kaori2008}, but requires a deeper understanding of how these properties emerge from the non-linear response of the complex gel structures.

Computational approaches and coarse grained numerical simulations are very effective tools to understand the flow, deformation, and microstructural evolution of particulate gels, providing new unique insight into the microstructural origin of the macroscopic behavior \cite{Bouzid:2018BookChap,zia14,jamali2017microstructural,varga2018normal}. In this work, we have combined coarse grained numerical simulations of model particulate gels, made of one and two components, with Large Amplitude Oscillatory Shear (LAOS) numerical tests. In particular, we have investigated the non-linear rheological response of an interesting set of model gel materials, which can have quite different rheological properties via changing the particle content or adding a second component that can further modify the mechanical properties. 

In investigations of the rheological behavior of gels, it is common to use the dynamic storage ($G^\prime(\omega)$) and loss ($G^{\prime \prime}(\omega)$) moduli to characterize the changes in viscoelasticity in the gels under different conditions. This is accomplished by considering the components of the stress which are in-phase with an applied shear strain and shear rate, respectively. As originally defined, the dynamic moduli in these tests are actually average energetic terms\cite{tschoegl89}: $G^\prime(\omega)$ is proportional to the average energy stored per cycle and and $G^{\prime \prime}(\omega)$ is proportional to the average rate of energy dissipation. In the non-linear regime, the traditional dynamic moduli have a few potential drawbacks. By definition, the non-linear properties of the material are expected to vary significantly depending on the deformation applied. While the energetic definitions for the moduli do hold outside the linear regime\cite{tschoegl89,donley20}, the moduli no longer capture the full extent of the non-linear behavior of the material due to the fact that they are average measures. Additionally, the need for a full period to define the moduli limits the time-resolution accessible with them to one point per period, or $\Delta t = 1/\omega$, so it is difficult to discern whether properties evolve on shorter timescales. The sequence of physical processes (SPP) framework developed in \cite{rogers11,rogers12_jor-a,rogers12_jor-b,rogers17} specifically utilizes the partial derivatives of the stress response of a material with respect to strain and rate to define time-dependent moduli ($G'_t$ and $G''_t$), and it has been shown to be quite versatile in understanding the nonlinear responses of a range of experimental materials \cite{rogers11,armstrong20,clarke21,donley19_jnnfm,donley19_ra,erturk22,korculanin21,lee19_karj,poggi22}, numerical simulations \cite{park20}, and theoretical models\cite{rogers12_jor-b,park18}.

We have applied the SPP analysis framework to our coarse grained numerical simulations of LAOS in the model particulate gels. By combining the SPP analysis originally designed for experiments with microscopic information on structure and dynamics that only numerical simulations can provide, we demonstrate how deeper understanding of the rheological response can be obtained, which then suggests new directions for material design. 

The paper is organized as follows. In section \ref{Methods} we describe the numerical model and the simulations performed, including the microscopic analysis. The same section also contains an overview of the SPP analysis and information on its implementation for the numerical simulations. Section \ref{sec:1comp} discusses the non-linear rheology of one-component gels, combining the results of the SPP analysis with the information on the microscopic quantities. In section \ref{sec:2comp} we analyze instead the behavior of two-component gels with different relative amount of the two components. Each of these sections, which contain extended discussion of the results and their implications, end with a short summary of the main findings. We then outline some conclusions and ideas for future work in section \ref{sec:conclu}.

\section{\label{Methods} Methods}
\subsection{\label{Model} Numerical Model}
We use a model particulate gel consisting of self-assembling units (particles) that interact via a short-range attraction, $U_2$ and a three body term, $U_3$ which limits the bond angles and introduces a bending stiffness \cite{bantawa21,bouzid18,colombo14}. Molecular Dynamics (MD) simulations are implemented in a system of $N$ particles with position vectors \{$\mathbf{r}_1,..., \mathbf{r}_N$\} and interacting with the potential energy:
\begin{equation}
U\left(\mathbf{r}_1,..., \mathbf{r}_N\right)=\epsilon \left[ \sum_{i>j}U_2\left(\frac{\textbf{ r}_{ij}}{d}\right)+\sum_i\sum_{j>k}^{j,k\neq i}U_3\left(\frac{\textbf{r}_{ij}}{d},\frac{\textbf{r}_{ik}}{d}\right)\right]
\label{Potential}
\end{equation}
where $\textbf{r}_{ij}=\textbf{r}_j-\textbf{r}_i$, $\epsilon$ is the depth of the attractive well $U_2$ and sets the energy scale, and $d$ is the particle diameter, representing the unit length scale. In typical colloidal systems, $d$ corresponds approximately to the range $d \simeq10$ to $100$~nm and $\epsilon \simeq 10$ to $100$~$k_{B}T$, where $k_B$ is the Boltzman constant and T, typically room temperature. The two-body term $U_2$ in Eq. \eqref{Potential} is a Lennard-Jones (LJ) like potential, and is a combination of a repulsive core and a narrow attractive well. For particles separated by a distance $r$ (here and in the following, distance is expressed in units of $d$), it is written in the form:
\indent \begin{equation}\label{Two-body term}
U_2(r) = A \left ( a{r^{-18}} - {r^{-16}}  \right ), 
\end{equation}
for computational convenience. The exponents have been chosen to produce a short range attractive well ($\sim1.3d$).

In real gels, the roughness of the particle surfaces or the irregular shapes of the aggregates in contact can result in a significant hindrance of the relative particle motion upon aggregation, hence limiting the local coordination of particle in the gels and introducing bending stiffness. There is evidence of these phenomena from experiments, showing that local coordination of particulate gels can be limited to 2-4 contacts and that interparticle bonds can resist finite torques \cite{Campbell2005,Pantina2006,Dibble2008,Whitaker2019,Bonacci2020}. We have designed the a three-body term $U_3$ in Eq. \eqref{Potential} to introduce the energy costs associated with the constraints imposed by the nature of the particle and aggregates contacts. This term provides bending rigidity to inter-particle bonds ${\textbf{r}}$ and $\textbf{r}^{\prime}$ departing from the same particle. The functional form of this term has been implemented, again for computational efficiency, as:
\indent \begin{equation}\label{Three-body term}
 U_3(\textbf{r},\textbf{r}^\prime) =  B\Lambda(r)\Lambda(r^\prime)\exp\Bigg [-\bigg(\frac{\textbf{r} \cdot \textbf{r}^\prime} {rr^\prime}-\cos\overline{\theta}\bigg)^2 \bigg / w^2 \Bigg]
\end{equation}
where $B$, $\overline{\theta}$ and $w$ are dimensionless parameters, and the radial modulation function $\Lambda(r)$ decays smoothly as,
\begin{equation}\label{lambda}
\Lambda(r)= r^{-10}\left[1-(r/2)^{10}\right]^2 {\mathcal{H}}(2-r)
\end{equation}
where ${\mathcal{H}}$ is the Heaviside function.
As described in previous works \cite{bantawa21,bouzid18,colombo14}, a persistent gel network can be obtained with the following set of potential parameters: $A=6.27$, $a=0.85$, $B=67.27$, $\theta=65^{\circ}$ and $w=0.3$. With this choice of parameters, the resulting gel structures are thin, space filling networks where particles have coordination numbers $z$ mostly $2$ or $3$. For a detailed discussion of the parameter choices, see Bantawa et al.\cite{bantawa21} and references therein. 

For the binary gel networks, we use the approach described in Vereroudakis et al.\cite{vereroudakis20}. In these simulations, each gel component is composed of self-assembling units that interact with potential described by Eq. \eqref{Potential}, with the same potential parameters used for the one component gels, except the parameter $A$ in Eq. \eqref{Two-body term} which is varied between the two components. The choice $A=6.27$ allows us to obtain a percolated network of semi-flexible fibers for component 1, as just discussed, while we model the second component by keeping the other parameters constant and changing instead $A=0.5$, which changes the depth of the potential well to $0.1\epsilon$. This choice makes the particles in component 2 self-assemble into small aggregates between which bonds can break and reform easily, and do not form, on their own, a spanning persistent network. For the interaction between the different components we use We choose $A=6.27$, resulting in a strong affinity of component 2 with component 1 in the mixtures.

\subsection{Simulation method and gel preparation}

\begin{figure*}
\includegraphics{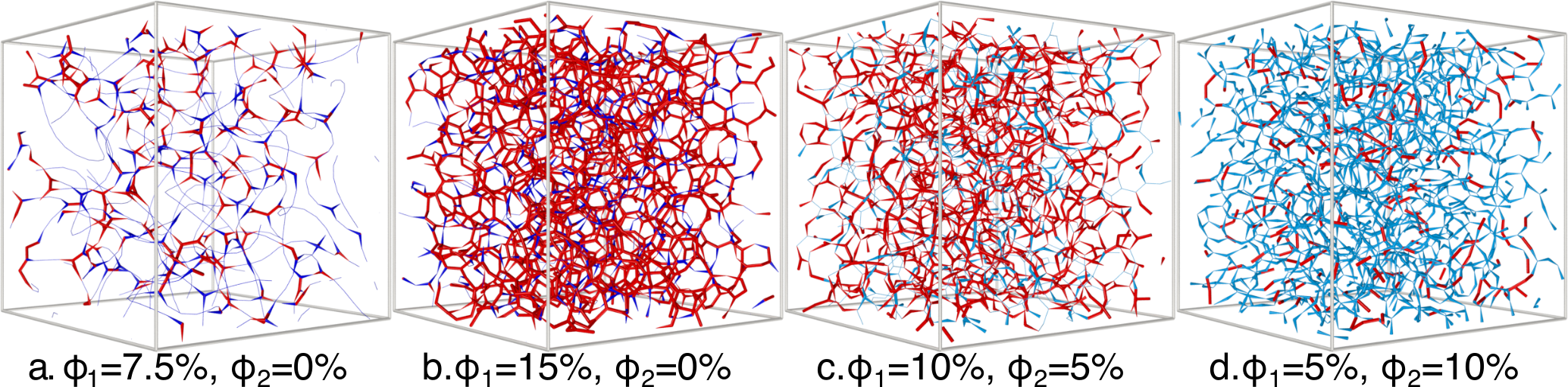}
\caption{\label{fig:struct} Snapshots of a portion of \color{black}the initial structure in four out of the six gels investigated in this study: a) the $\phi=7.5\%$ 1-component gel, b) the $\phi=15\%$ 1-component gel, c) the $\phi_1=10\%$, $\phi_2=5\%$ 2-component gel, and d) the $\phi_1=5\%$, $\phi_2=10\%$ 2-component gel. The configurations in a-b) show the effect of a change in volume fraction on a fixed composition (100\% component 1), with dark blue portions indicating a coordination number of 2 and red portions indicating a coordination number of 3. Meanwhile, c-d) show show the effect of a change in composition at a fixed total volume fraction ($\phi_{total}=15\%$, same as b)), with red indicating component 1 and light blue indicating component 2. In all cases the line thickness denotes the coordination number, with thin lines showing strands(coordination number of 2) and thick lines showing branching points (coordination number of 3).}
\end{figure*}

We perform MD simulations using the open-source LAMMPS software~\cite{plimpton95}, appropriately modified to incorporate the interaction potential in Eq. \eqref{Potential}. We report results with $N = 16384-32768$ particles in a cubic box of size $L$ and number density $N/L^3$ which corresponds to an approximate solid volume fraction $\phi =\frac{N \pi d^3/6}{(Ld)^3}$. Here we discuss 1-component gels with total number of particles fixed at $N=16384$ and different volume fractions: $\phi=7.5\%, 10\%, 15\%$. The 2-component gels consist $N=24576$ for total volume fraction $\phi_{total}=15\%$ and $N=32768$ for $\phi_{total}=20\%$.  We use periodic boundary conditions and solve the equations of motion with the interactions described in section \ref{Model} and a time step $\delta t = 0.005\tau_0$ ($\tau_0 = \sqrt{md^2/\epsilon}$ is the usual MD time unit).

\subsubsection{1-component network} \label{gel:prep}
We prepare the initial single component gel configurations by following the protocol described by Bantawa et al.~\cite{bantawa21}. The particles interact with the potential described by Eq.~\eqref{Potential} and the parameters defined in Section \ref{Model}. The preparation of the initial configurations is performed in two parts. First, we start from particles placed randomly in a cubic simulation box of size $L$, initially equilibrated at $k_BT/\epsilon=0.5$ and cooled down slowly, using a Nose-Hoover (NH) thermostat in an equilibrium NVT simulation, to $k_{B}T/\epsilon =0.05$ so that system spontaneously self-assembles into a network of strands (particles that have coordination number $z=2$) connected by branching points ($z=3$). As already done in previous studies \cite{bouzid2017elastically,bouzid18}, we use a cooling rate of $\Gamma \approx 10^{-5} \epsilon /k_B\tau_0$, for which the microstates obtained do not significantly depend on the dynamics used. Therefore, in this part of the gel preparation, the simple NVT MD can be used to reduce the simulation time with respect to the more physically meaningful, but computationally slower, Langevin dynamics. We finally further equilibrate the system at $k_{B}T/\epsilon = 0.05$ with the NH thermostat for additional $2\cdot10^4$ MD steps. 

In the second part of the gel preparation, we employ a damped dynamics to relax the configuration, obtained at a finite temperature, to a local minimum that more likely corresponds to a mechanically stable configuration. This is achieved by withdrawing the kinetic energy of the system to $\sim 10^{-10}$ of its initial value with an overdamped dissipative dynamics:
\begin{equation}
m\frac{d^2\textbf{r}_i}{dt^2}=-\nabla_{\textbf{r}_i}U-\zeta\frac{d\textbf{r}_i}{dt},
\end{equation}
where $m$ is the mass of each particle and $\zeta$ represents the drag coefficient of the surrounding solvent. The relaxation of the configuration could be, in principle, also carried out by performing the total energy minimization of the system with the conjugate gradient algorithm. However, previous studies \cite{colombo14,bouzid18,bantawa21} showed that this procedure is more efficient with very soft gels and we have therefore used this damped dynamics for the energy minimization in all simulations. Overall, the gel preparation protocol was designed for computational efficiency and the gel structures emerge from the balance between entropy and interaction energy, without imposing a specific aggregation kinetics. Examples of the structures formed by this system are shown in Fig.\ref{fig:struct}a-b for the $\phi=7.5\%$ and $\phi=15\%$ gel cases. At the larger volume fraction the density of crosslinks is significantly higher, whereas there are longer strands in the less concentrated gel, whose microstructure is more heterogeneous in general.

\subsubsection{2-component network}

The preparation of the initial configurations for the 2-component gels utilizes the same two step procedure described for 1-component gel in Section \ref{gel:prep}. In the present study, we consider three cases with different ratios of component-1 and component-2 : 1) $\phi_1=10\%$ and $\phi_2=5\%$ (i.e. $\phi_{total}=15\%$, component ratio 2:1); 2) $\phi_1=10\%$ and $\phi_2=10\%$ ($\phi_{total}=20\%$, 1:1); and 3) $\phi_1=5\%$ and $\phi_2=10\%$ ($\phi_{total}=15\%$, 1:2). The two $\phi_{total}=15\%$ cases are shown in Fig.\ref{fig:struct}c and Fig.\ref{fig:struct}d, respectively, where the different colors distinguish the two components.

\subsection{Microstructural quantities}
In the simulations we can characterize the changes in the microstructure due to different amount of deformations and connect them to the rheological response. The quantities we investigate here are the Shear-induced Anisotropy (S), the statistics of bonds broken and new bonds formed, and the non-affine displacements (NAD) during deformation. Below we describe these quantities.

\subsubsection{Shear-induced Anisotropy (S)} \label{S}
Following Bouzid et al.\cite{bouzid18}, the local alignment of bonds or strands upon deformation is measured in terms of a nematic tensor $Q_{ab}$:
\begin{equation}
 Q_{ab} = \frac{1}{2} \langle 3 {\textbf{n}}_a {\textbf{n}}_b - \delta_{ab} \rangle .
 \label{Q-tensor}
\end{equation}
where ${\textbf{n}}_a$ and ${\textbf{n}}_b$ represent the unit vectors corresponding to the orientation of neighboring bonds $a$ and $b$, and $\langle ... \rangle$ indicates an average over all bonds. The largest positive eigenvalue of $Q$ is represented by a scalar order parameter $S$ which is the measure of the average anisotropy in the bond or strands orientation. $S=0$ corresponds to a random orientation and $S=1$ to the case when all bonds are fully aligned along the same direction. This quantity is measured during the rheological oscillation cycles described in the next section. In the related plots, we additionally average $S$ over $2$ successive cycles of the oscillation, and apply a $5$-point moving average \cite{Kennybook} to smooth the signals.

\subsubsection{Bond statistics}
Two particles are considered bonded if they are separated by a distance $\leq 1.3d$ (range of two-body interaction) \cite{colombo14}. We then monitor the evolution of the number of new bonds formed or bonds broken between successive time steps. This measure allows us to identify the plastic processes, occuring within the parts of the microstructure and due to local rearrangements in response to the applied deformation. In the related plots, we average these quantities over $4$ successive cycles of the rheological oscillations, and apply a $5$-point moving average \cite{Kennybook} to smooth the signal.

\subsubsection{Non-affine displacements (NAD)}
In the simulations, we consider that the shear deformation (time dependent strain $\gamma(t)$) imposed to an initial particle configuration $\{\textbf{r}_i\}$ includes an instantaneous affine deformation $\Gamma_{\gamma}$ in the $xy$ plane of simple shear to all particles \cite{colombo14}, so that the deformed configuration is $\{\textbf{r}^\prime_i\}$ with the following transformation:
\begin{equation}\label{affine}
  \textbf{r}_i^\prime = \Gamma_{\gamma}\textbf{r}_i = 
  \begin{pmatrix}
    1 & \gamma & 0 \\
    0 & 1 & 0 \\
    0 & 0 & 1   
  \end{pmatrix}\textbf{r}_i
\end{equation}
We then compute the non-affine displacement (NAD) as the displacement of the actual particle position $\{\textbf{r}_i^\mathrm{actual}\}$ w.r.t. the particle position due to the affine transformation as follows:
\begin{equation}\label{NAD}
    \mathrm{NAD}_i=||\textbf{r}_i^\mathrm{actual}-\Gamma_{\gamma}\textbf{r}_i||
\end{equation}
This NAD measure can be used to characterize the overall and local amount of non-affine motion at any given time, as the oscillatory strain is being imposed during the rheological tests.

\subsection{\label{rheology} Rheology}

\subsubsection{\label{rheo_measurements} Measurements}

The rheological response is measured by imposing an oscillatory strain signal $\gamma(t)=\gamma_0 \sin \omega t$ in the $xy$-plane of the simulation box through
\begin{equation}\label{oscillatory}
 m\frac{d^2{\textbf{r}}_i}{dt^2}=-\nabla_{{\textbf{r}}_i}U - \zeta\left(\frac{d{\textbf{r}}_i}{dt}-\dot{\gamma}(t)y_i{\hat{\textbf{x}}}\right) 
\end{equation}
while updating the Lees-Edwards boundary conditions at every time step. Here $\gamma_0$ is the strain amplitude and $\hat{\textbf{x}}$ denotes the unit vector in the $x$-direction. In all the rheological measurements we used a Stokes-like drag with $m/\zeta=0.5\tau_0$, having verified that the results do not change qualitatively with further decreasing the $m/\zeta$ ratio.
 
To systematically study how the gels' nonlinear rheological responses depend on their microstructures, we perform large amplitude oscillatory shear (LAOS) using a rate-controlled deformation:
\begin{equation}\label{osc_def}
\dot{\gamma}(t)=\gamma_0\omega cos(\omega t).
\end{equation}
We specifically chose to perform an amplitude sweep by varying the amplitude over the range $0.001 < \gamma_0 < 10$ strain units and holding the frequency of oscillation at $\omega$ = 0.0025$\tau_0^{-1}$. This range of strain amplitudes spans from the linear viscoelastic regime through the point where deformations are large enough to yield the material. It is worth noting that, in these simulations, every amplitude in the amplitude sweep corresponds to a separate rheological test that starts from the same unperturbed gel configuration. Hence, differently from typical LAOS strain sweeps in experiments, we do not superimpose the effects of the shear history due to different amplitudes, in an attempt to focus on the consequences of the nonlinear deformations.

Under applied deformation, we compute the instantaneous shear stress $\sigma_{xy} (t)$. The stresses are computed from the interaction part of the global stress tensor using the standard virial equation \cite{thompson09} while neglecting other contributions (kinetic and viscous terms) as in previous studies \cite{colombo14,Bouzid:2018BookChap}:
\begin{equation}\label{Virial}
\sigma_{\alpha\beta} = \frac{1}{L^3} \sum_{i=1}^N\frac{\partial U}{\partial r_i^\alpha}r_i^\beta,
\end{equation}
where $\alpha$ and $\beta$ denote the Cartesian components $\{x,y,z\}$. This choice allows us to focus on the structural contribution to the gels viscoelasticity.

All results discussed here are shown once the stress response has reached steady alternance, which takes 4 cycles for each of these gel samples, as verified by carefully checking the superposition of Lissajous plots\cite{donley19_jnnfm}. For microstructural quantities averaged over multiple cycles, we consider cycles after the steady alternance has been reached. For the numerical calculations based on the finite time series of strain and stress data, we use Fourier domain filtering with a finite number of odd harmonics to create an analytical reconstruction of the signal. The exact number of harmonics is varied for each oscillation case such that the reconstruction contains as many harmonics as are above the noise floor of the signal.

\subsubsection{\label{rheo_moduli} Dynamic Moduli}

The complex viscoelastic modulus $G^*(\omega)$ is obtained from the Fourier transforms of the stress output ${\tilde{\sigma}}(\omega)$ and the strain input ${\tilde{\gamma}}(\omega)$ signals, as $G^*(\omega)={\tilde{\sigma}}(\omega)/{\tilde{\gamma}}(\omega)$, from which we compute the storage modulus $G^\prime(\omega)$ and the loss modulus $G^{\prime \prime}(\omega)$ defined respectively as the real and imaginary part of $G^*(\omega)$. Both oscillatory shear and steady shear measurements \cite{colombo14,bouzid18} indicate that the dynamic moduli are linear at small strain amplitudes ($\gamma_0$)  while the response becomes nonlinear at large $\gamma_0$ values.

\subsubsection{Understanding LAOS}

A number of different analysis frameworks have been used to characterize large-amplitude oscillatory shear (LAOS) deformation. These mainly fall into three categories: 1) methods that extend the linear regime Fourier analysis to higher harmonics\cite{klein07,laurati14,wilhelm02}, 2) methods which geometrically decompose the LAOS response into contributions dependent on strain and strain rate\cite{cho05,dimitriou13,ewoldt08,ewoldt10}, and 3) derivative-based methods which use the trajectory of the response to define time-resolved moduli\cite{rogers11,rogers12_jor-a, rogers12_jor-b,rogers17}.

The harmonic based techniques do capture all of the non-linearity of the LAOS data, but aside from the third harmonic, no specific meaning has been successfully assigned to each of these values\cite{ewoldt08}, making comparisons challenging at large amplitudes. These methods also suffer the drawback of requiring a full period to be defined, thus limiting their ability to capture any changes which may occur within a period\cite{donley19_jnnfm,donley19_ra}. The analytical decomposition techniques do allow for increased resolution within a single period, but they rely on symmetry assumptions which are not confirmed to work for all materials, and have been shown to be incorrect in specific cases\cite{ewoldt08,mermetguyennet14,rogers11,rogers17}.

Of the three methods, those based on derivatives of the material properties are the most robust: they offer resolution within a single period limited only by the density of data points, and make very few assumptions about the specific properties of the material being investigated. The most prominent of these methods is the Sequence of Physical Processes (SPP) framework of Rogers and collaborators \cite{rogers11,rogers12_jor-a,rogers12_jor-b,rogers17}.

\subsubsection{\label{rheo_spp} Sequence of Physical Processes (SPP) Framework}

The SPP framework assumes that all changes in the material properties within a cycle of deformation are due to changes in the values of time-dependent moduli which maintain an instantaneous linear relationship\cite{rogers17}:
\begin{equation}\label{eqn_spp}
\sigma(t) = G'_t(t)\gamma(t)+\frac{G''_t(t)}{\omega}\dot{\gamma}(t)+\sigma^d(t),
\end{equation}
where $G'_t$ and $G''_t$ are the time-dependent dynamic moduli, and $\sigma^d$ is a vertical offset in the stress\cite{rogers17}. The values of the time-dependent moduli are determined from the partial derivatives of the transient material response:
\begin{eqnarray}
 G'_t (t) = \frac{\partial \sigma (t)}{\partial \gamma (t)}
 \label{eqn_gpt}\\
 G''_t (t) = \omega \frac{\partial \sigma (t)}{\partial \dot{\gamma} (t)} .
 \label{eqn_gppt}
\end{eqnarray}

The changes in these moduli can be used to track transitions that occur within a single period of oscillation. Specifically, $G'_t(t)$ tracks the elastic behavior of the material, while $G''_t(t)$ tracks the viscous behavior. Increases in the moduli show in instantaneous stiffening ($G'_t$) or thickening ($G''_t$) of the material, while decreases show instantaneous softening ($G'_t$) or thinning ($G''_t$)\cite{rogers17,donley19_ra,choi19}.

It is worth noting that the instantaneous linear relationship of the moduli in Eq.~\eqref{eqn_spp} is unable to distinguish instantaneously non-linear behaviors. For example, prior studies \cite{rogers12_jor-b} have shown that constitutive models which assume instantaneous shear-thinning register as having variable elasticity under the SPP framework. While not seen in prior studies, it is possible, by similar reasoning, that instantaeous non-linear elasticity may manifest as variable viscosity within this framework.

\section{\label{sec:1comp} Results for 1-Component Gels}
\subsection{\label{1c_rheo} Rheology}
\subsubsection{\label{1c_laos} Oscillatory Rheology}

\begin{figure*}
\includegraphics{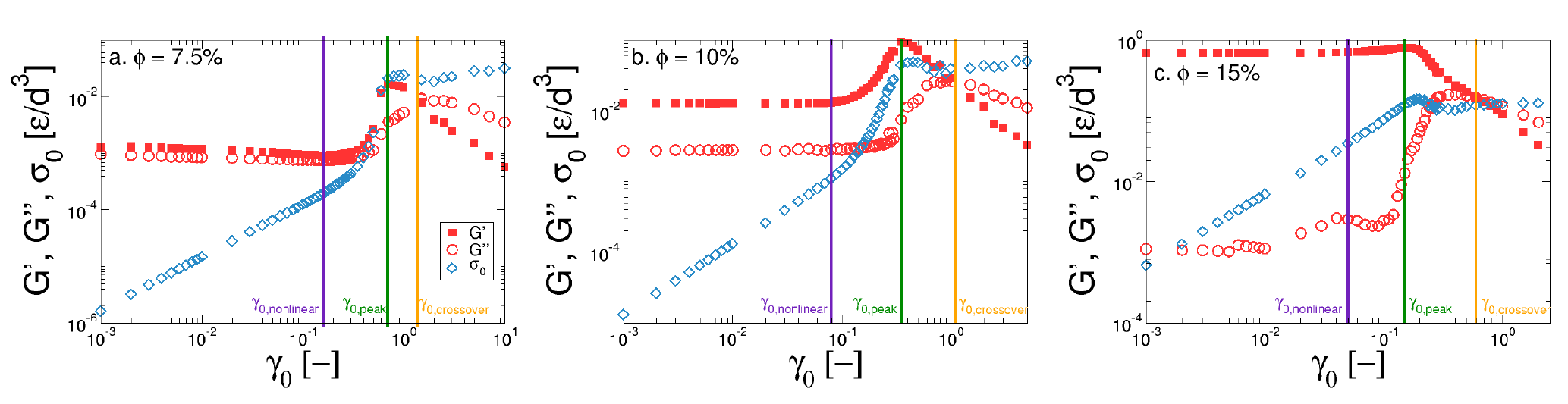}
\caption{\label{fig:asweep} Amplitude sweep results for the three 1-component gel concentrations: a) $\phi=7.5\%$, b) $\phi=10\%$, and c) $\phi=15\%$. The vertical lines denote threshold amplitudes used for comparing the rheology and microstructure of the different gels: the first amplitude outside the linear regime (purple), the amplitude at the peak of $G'$ (green) and the amplitude of the moduli crossover (orange).}
\end{figure*}

The amplitude sweeps for the single component gels are shown in Fig.~\ref{fig:asweep}. At each of the three volume fractions, the responses exhibit three regimes: 1) a linear regime at small amplitudes with $G'>G''$, 2) a region of non-linear elasticity at intermediate amplitudes distinguished by the overshoot in $G'$ \cite{bouzid18,colombo14}, and 3) yielding behavior at large amplitudes as indicated by the overshoot in $G''$ and the crossover point in the moduli~\cite{colombo14,donley20}. This type of amplitude sweep behavior with overshoots in both $G'$ and $G''$, followed by a gradual drop, has been classified as strong strain overshoot behavior\cite{hyun11} and may be consistent with an elastic matrix which can dynamically yield. 

It should be noted that the stress amplitude ($\sigma_0$, blue circles in Fig.~\ref{fig:asweep}) shows signatures of these same three regions. In the linear regime, $\sigma_0$ increases linearly with strain amplitude ($\gamma_0$), with a prefactor of $G^*=\sqrt{G'^2+G''^2}$. The slope increases with the onset of non-linear elasticity, before $\sigma_0$ overshoots and then plateaus at large amplitudes due to the yielding of the gel. 

Over the three gels, the values of the linear regime elastic modulus ($G'_{linear}$) and the ratio of $G'_{linear}/G''_{linear}$ increase with increasing volume fraction. The strain amplitudes at which the material transitions out of the linear response regime, into the yielding and into flow, also decrease as a function of volume fraction. These findings may point to more constraints emerging in the microstructure, with increasing the volume fraction, due to an increase in gel branching points (see Fig. \ref{fig:struct}a,b) \cite{bouzid18}. We note that the relative size of the $G'$ overshoot decreases instead as volume fraction increases, while $G''_{linear}$ peaks in the $\phi=10\%$ case. In the following, we will combine the time-resolved rheological data with the microstructural analysis to rationalize these observations.

\subsubsection{\label{sec:spp_results} Time-resolved Rheology}

\begin{figure*}
\includegraphics{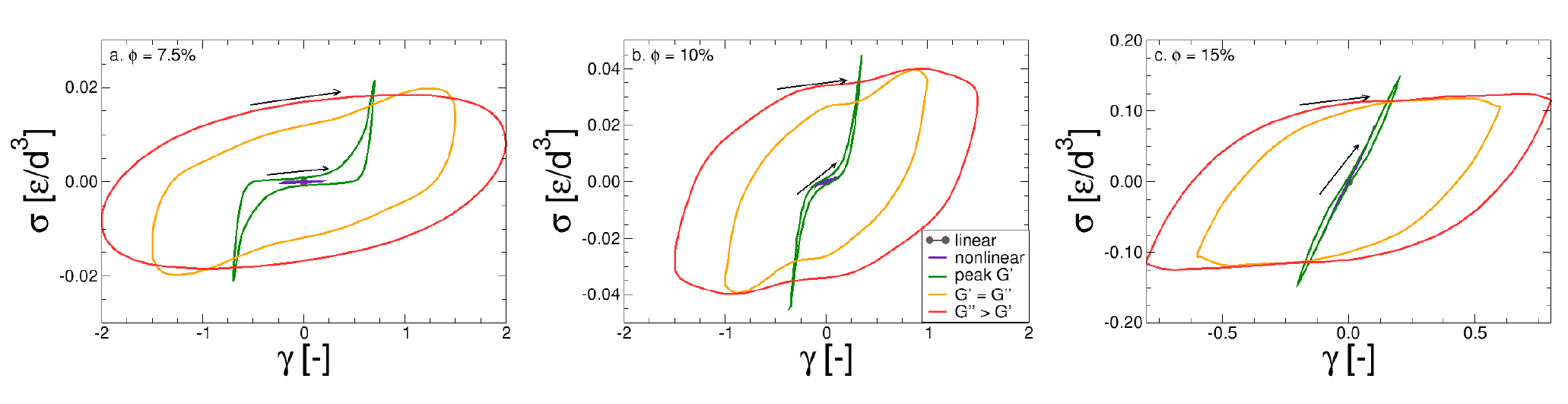}
\caption{\label{fig:laos} Elastic Lissajous curves of the time-resolved LAOS for the different 1-component gel concentrations: a) $\phi=7.5\%$, b) $\phi=10\%$, and c) $\phi=15\%$. The different colors refer to the key amplitudes referred to in the main text. The black arrows denote direction of curve throughout the oscillation.}
\end{figure*}

\begin{figure*}
\includegraphics{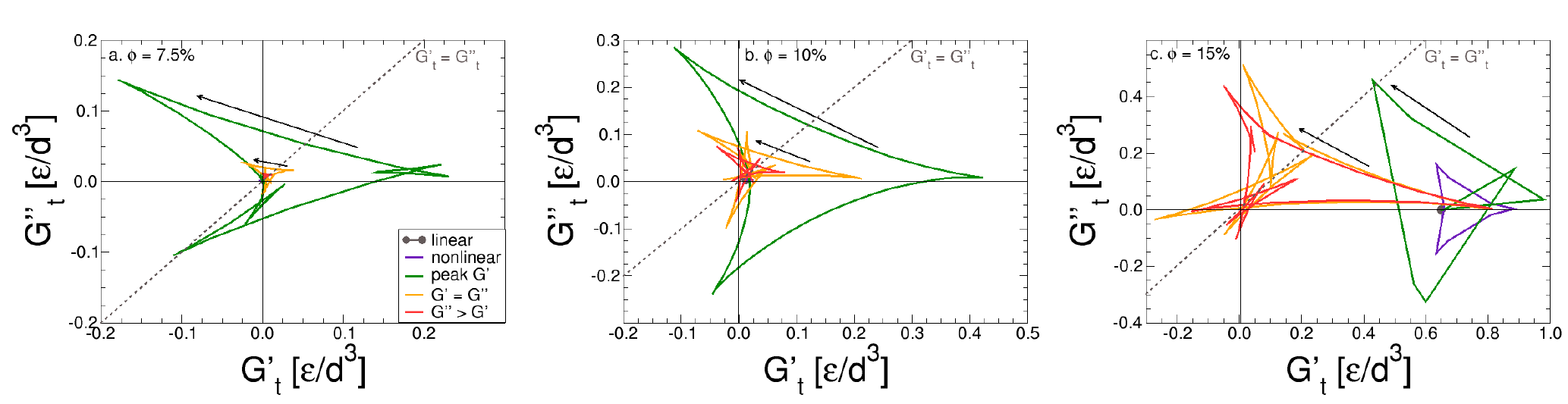}
\caption{\label{fig:spp} Cole-Cole plots of the time-resolved dynamic SPP moduli ($G^\prime_t(\omega,t)$ and $G^{\prime \prime}_t(\omega,t)$) for the different 1-component gel concentrations: a) $\phi=7.5\%$, b) $\phi=10\%$, and c) $\phi=15\%$. The different colors refer to the key amplitudes referred to in the main text. The black arrows denote direction of curve throughout the oscillation.}
\end{figure*}

As the gels appear to reach their respective regimes of non-linear elasticity and yielding behavior at different applied strain amplitudes, it makes more sense to compare them at points of similar phenomenology, rather than at a fixed amplitudes. To this end, we will use 5 threshold amplitudes to compare the time resolved rheological behavior and microstructural information: 1) the smallest strain amplitude tested (linear viscoelastic (LVE) regime, defined as any amplitude where the oscillatory behavior is completely described by a single harmonic), 2) the first amplitude outside the linear regime (transition from LVE to nonlinear elasticity), 3) the peak in $G'$ (transition from nonlinear elasticity to yielding), 4) the dynamic modulus crossover $G_X=G'(\gamma_X)=G''(\gamma_X)$ (transition from yielding to flow), and 5) a LAOS amplitude ~30$\%$ larger than the crossover point (flow behavior in large-amplitude limit). The specific values used for each of the three 1-component gel formulations are listed in Table \ref{tab:1comp}.

\begin{table}
\caption{\label{tab:1comp} Applied strain amplitudes used for rheology and structure comparisons between the 1-component gel formulations. All amplitudes are given in strain units.}
\begin{ruledtabular}
\begin{tabular}{llll}
 & $\phi$=7.5\%vol&$\phi$=10\%vol&$\phi$=15\%vol\\
\hline
linear & 0.001 & 0.001 & 0.001\\
non-linear & 0.20 & 0.10 & 0.10\\
peak $G'$ & 0.70 & 0.35 & 0.20\\
$G'$=$G''$ & 1.5 & 1.0 & 0.60\\
$G''$>$G'$ & 2.0 & 1.5 & 0.80\\
\end{tabular}
\end{ruledtabular}
\end{table}

The time-resolved rheology for each of the threshold amplitudes is shown in Fig.~\ref{fig:laos} as elastic Lissajous curves (plots of stress $\sigma$ vs. strain $\gamma$). In this representation, perfectly elastic materials are seen as straight lines through the origin, perfectly viscous materials appear as horizontal ellipses, and linear viscoelastic responses are tilted ellipses. Distorted (non-elliptical) elastic Lissajous curves are produced when a response is non-linear and/or changes with time.

For each volume fraction, at the smallest two threshold amplitudes (grey and purple curves corresponding respectively to the smallest amplitude tested in the linear regime and the end of that regime), the curves appear to be linear, suggesting a predominantly linear elastic behavior. For these smaller amplitudes, the Lissajous curves are tilted ellipses, although not clearly visible in the figure, therefore indicating linear viscoelasticity. At the peak in $G'$ (green curve), the slope of the curve increases drastically near the strain extrema for the $\phi$ = 7.5$\%$ (Fig.~\ref{fig:laos}a) and $\phi$ = 10$\%$ (Fig.~\ref{fig:laos}b) gels, suggesting a strong non-linear elasticity which depends on the applied strain. This is also seen at $\phi$ = 15$\%$ (Fig.~\ref{fig:laos}c), albeit to a much less pronounced extent. Hence the time-dependent rheological analysis point to the fact that the nonlinear elastic nature of the gels varies drastically and decreases with the gel volume fraction. Going back to the overall behavior of the time-averaged rheological response described in the previous section, these findings indicate that the decrease of the relative size of the $G'$ overshoot is associated to a decrease of the nonlinear elastic response. By the time the crossover point is reached (orange curve), the elastic Lissajous curves have opened significantly, indicating an increase in viscous dissipation. At the largest amplitudes, the $\phi$ = 7.5$\%$ gel appears to move toward a nearly elliptical response, suggesting a viscoelastic fluid behavior, possibly reaching a quasi-linear large amplitude regime for large enough amplitudes \cite{desouzamendes14}. The $\phi$ = 15$\%$ gel, on the other hand, trends toward a parallelogram-like shape which is characteristic of the LAOS behavior of yield stress fluids\cite{donley19_jnnfm}. The $\phi$ = 10$\%$ gel shows an intermediate signature with elements of both types of fluids. 

To track the time-dependent rheology within each period, we plot the time-resolved dynamic moduli $G'_t$ and $G''_t$ in Fig.~\ref{fig:spp}, displayed in the form of time-dependent Cole-Cole plots. In this representation, portions of the trajectory which travel parallel the x-axis ($G'_t$) show changes in the elasticity of the materials (i.e. an increase is stiffening while a decrease is softening), while portions traveling parallel to the y-axis ($G''_t$) show changes in the viscosity of the materials (i.e. an increase is thickening while a decrease is thinning). Finally, transitions across the $G'_t=G''_t$ line represents transitions between predominantly elastic and viscous states. Further discussion of how to read these plots can be found elsewhere \cite{donley19_ra,rogers17,choi19}.

In all cases, the linear regime moduli have a constant value throughout the period, which is exactly equal to the linear-regime value seen in the amplitude sweeps in Fig.~\ref{fig:asweep}, with $G'_t>G''_t$. As the amplitude increases into the non-linear regime, the fact that the material properties are changing as a result of the applied deformation manifests itself in  $G'_t$ and $G''_t$ varying significantly throughout the period. The trajectory of the moduli tends to form the shape of a deltoid in the Cole-Cole plots, with the center of mass of the curve being equal to the traditional modulus as seen in the amplitude sweep, and the area of the deltoid representing the extent of the non-linearity. Just outside the linear regime (purple curve, which is clearly visible only for $\phi=15\%$) this is small, but it grows substantially by the time the peak in $G'$ is reached for all volume fractions. At this stage, we see that the materials are already transitioning between predominantly solid-like and fluid-like behavior at different points in the period. This indicates that some dissipation process, such as viscous reorientation of the network possibly due to the breaking of parts of the microstructure, is instantaneously dominant over the elastic deformation for a portion of the period, but the elastic response still dominates on average, suggesting that an overall rigid network is still present. There are also regions in these plots with instantaneous negative values of the moduli, which suggests that a process such as elastic recoil (in the case of $G'_t<0$) or viscous backflow (in the case of $G''_t<0$) may be occurring. These phenomena are possible, so long as the average value of the corresponding modulus remain positive. In a few cases (predominantly at larger amplitudes or higher volume fractions), the deltoids can be seen to exhibit smaller loops in some portion of the trajectory. These are most likely artefacts resulting from the finite number of harmonics used in the reconstruction of the stress and strain data, and as such will not be looked at in detail in this analysis. 

In the Cole-Cole plots, it is at the peak in $G'$ where we first begin to see substantial differences between the different volume fractions. Both the $\phi$ = 7.5$\%$ and $\phi$ = 10$\%$ samples are more viscous than the $\phi$ = 15$\%$ gel, which barely crosses the $G'_t=G''_t$ and does not show any signature of elastic recoil at this point. By the time the crossover point is reached (orange curve), the differences are even more apparent. The size of the $\phi$ = 7.5$\%$ gel's deltoid decreases dramatically, indicating that, while the macroscopic behavior is relatively constant with time, it is clearly distinct from the one in the linear regime \cite{desouzamendes14}, also consistent with the behavior detected by the corresponding elastic Lissajous curve (see Fig.~\ref{fig:laos}). At the other extreme, the $\phi$ = 15$\%$ gel maintains its strong linear regime elastic value for a very small portion of the period, but spends a large portion of the oscillation in a predominantly viscous state, a response which is characteristic of a yield stress fluid close to or above the yielding point \cite{donley19_jnnfm,donley19_ra}, and suggesting that the decrease in the nonlinear elastic response in this sample may be accompanied by a pronounced increase of plasticity in the microstructure. The $\phi$ = 10$\%$ sample is again intermediate between the other two gels, with the elasticity much reduced but still maintaining some of the yield stress fluid  character.

With respect to variation of $G''_t$ across the different volume fractions, we note that some of it may be due to the instantaneous linear assumption of the SPP framework (Eq.~\eqref{eqn_spp})\cite{rogers12_jor-b}
breaking down, rather than a true variation in the viscous behavior of the material. This may be particularly important in the cases where the nonlinear elasticity is pronounced (as indicated by the elastic Lissajous curves): the fact that both $G'_t$ and $G''_t$ amplitudes appear to show some level of symmetry around the average $G''_t$ value, particularly in the $\phi$ = 7.5$\%$ and $\phi$ = 10$\%$ samples, suggests that both non-linear elasticity and transient viscosity are contributing to the time-resolved behavior.

\subsection{\label{1c_structure} Rheo-structural Characterization}

As discussed in the previous section, the rheology of the particulate gels studied here demonstrates two distinct non-linear effects: non-linear elasticity and reversible yielding, whose predominance in the rheological response seems to vary non-monotonically with the gel volume fraction. To understand the microstructural origin of these non-linear effects, we will look at three time-resolved microstructural measures. The presence of non-linear elasticity may be explained, in these gels, by the alignment and stretching out of the bond network \cite{colombo14,bouzid18}, indicating that shear-induced anisotropy will be useful to investigate. If the gel undergoes a reversible yielding transition the statistics of the breakage and formation of bonds will likely be insightful. Finally, the amount of non-affine motion of the gel particles will enable us to detect the possible microscopic effects of the yielding transition on the flow and transient structure of the materials.

\subsubsection{\label{sec:shinani} Shear-induced Anisotropy}

\begin{figure*}
\includegraphics{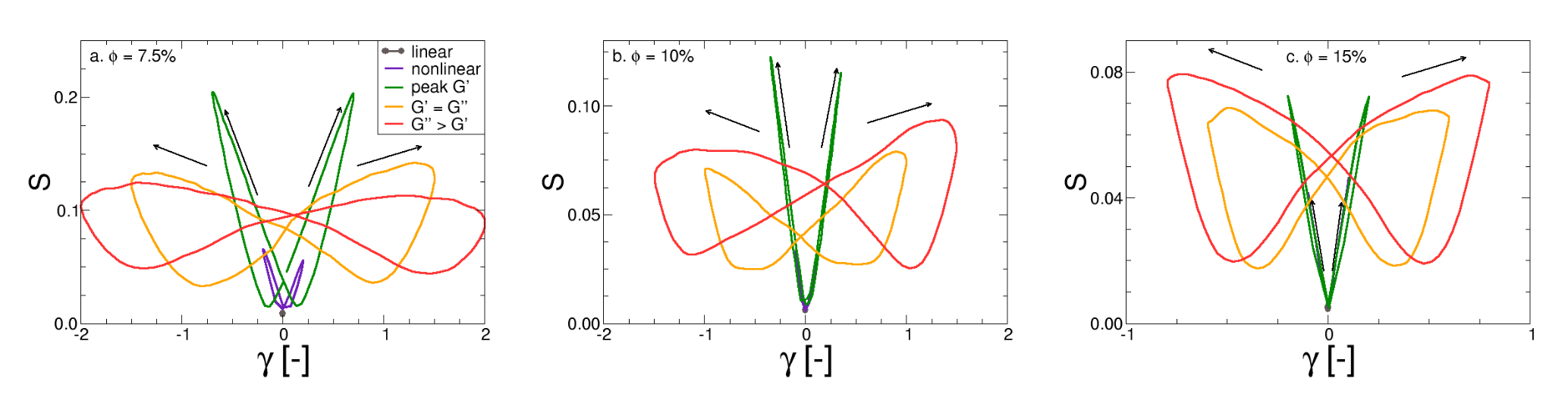}
\caption{\label{fig:nemord} Comparison of the nematic order parameter $S$ with the strain $\gamma$ under LAOS deformation for the different 1 component gel concentrations: a) $\phi=7.5\%$, b) $\phi=10\%$, and c) $\phi=15\%$. The different colors refer to the key amplitudes referred to in the main text. The black arrows denote direction of curve throughout the oscillation.}
\end{figure*}

To characterize the anisotropy that the gel microstructures can acquire under shear, we use the scalar order parameter $S$ defined in Section \ref{S}, and shown in Fig.\ref{fig:nemord} as a function of the applied strain during the LAOS tests. For amplitudes at or below the peak in $G'$, $S$ appears to scale proportionately to the strain, with the specific scaling dependent on the volume fraction. This relationship is linear for the $\phi$ = 10$\%$ and $\phi$ = 15$\%$ gels, while there is a pronounced hysteresis in the $\phi$ = 7.5$\%$ case. As the applied strain amplitude increases toward the crossover point, all curves shift to a butterfly-like shape.

The butterfly pattern in $S$ indicates that the structure initially aligned under shear leads to a distinct path through which the alignment can be modified when the strain direction is inverted over the cycle. The new path could be associated to a change of the bond network (through breaking or formation) under shear or to the possibility of large non-affine motion that can occur even without major changes of the bond network in very soft structures \cite{colombo14,bouzid18}. We note that, across all gels, the maximum alignment below the yielding point decreases with increasing volume fraction, while the ratio of alignment before and after yielding increases. These observations, combined with the dependence, discussed above, of the nonlinear elasticity on the gels volume fraction, may already suggest that the alignment quantified by $S$ is increasingly due to changes in the bond network upon increasing the volume fraction and with decreasing nonlinear elasticity of the microstructures. The investigation of the statistics of bond breaking and formation in the next section will help us confirm this interpretation. 

The butterfly patterns we find here are reminiscent of similar patterns seen in experiments in colloidal gel  \cite{colombo17,hoekstra05,kim14,masschaele11,park17,park20,reddy12,varadan01,vermant05}, as well as other self-assembled microstructures \cite{lee19_prl,rogers12_sm}, hence understanding  their microscopic origin could also help us connect the behavior of the gels investigated here to those experimental systems. It should be noted, however, that in most case experiments report a 2D projection of the structural alignment (i.e. flow-gradient, flow-vorticity, or gradient-vorticity), whereas here we are only evaluating the overall alignment in all possible directions.  

The fact that the maximum value of $S$ occurs at the peak in $G'$ (green curves) supports the idea that the scalar order parameter $S$ is predominantly coupled to the elastic deformation of the samples. At small amplitudes, the nearly linear relationship between $S$ and $\gamma$ suggests that nearly all of the strain results in elastic deformation. The emergence of the butterfly patterns above yielding, instead, suggests that significant microstructural changes have divorced the maximal elastic deformation from the total strain. Hence, taken together, these findings indicate that the overall amount of alignment $S$ could be directly related to the recoverable strain in the system \cite{lee19_prl,donley20}, rather than the total strain, consistent with experimental observations in self-assembled wormlike micelle solutions\cite{lee19_prl}. In simple yield stress fluids, the majority of strain below yielding is typically recoverable, while the total strain is much closer to the unrecoverable strain above yielding \cite{donley20}. Similarly, in our gels $S$ correlates well with the total strain at small amplitudes yet it does not at large amplitudes. For the $\phi$ = 7.5$\%$ gel, where the decoupling between $S$ and the total strain starts below yielding, the less connected microstructure more prone to non-affine motion may provide a source of unrecoverable strain already below yielding.

\subsubsection{\label{sec:bbf} Bond Breakage/Formation}

\begin{figure*}
\includegraphics{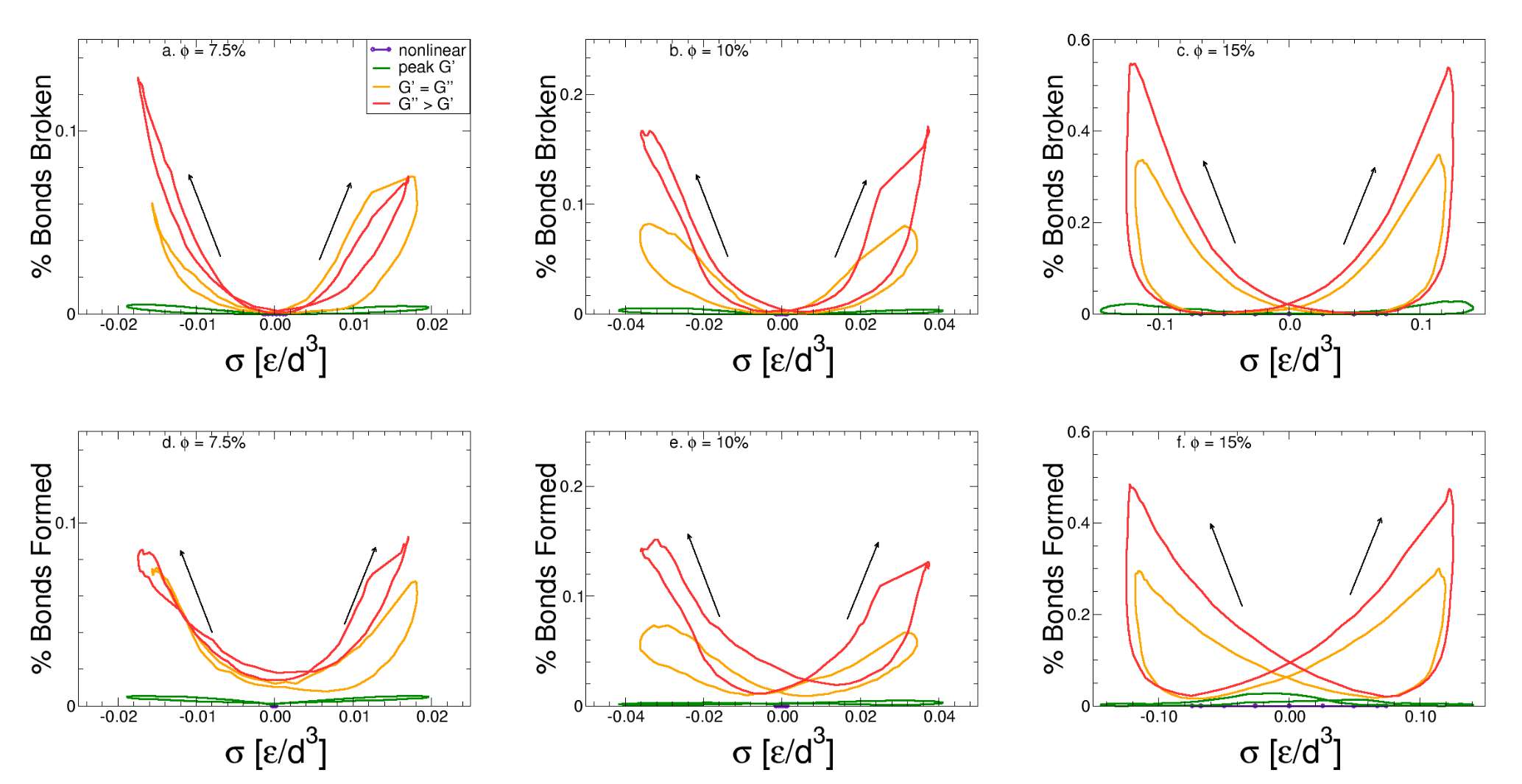}
\caption{\label{fig:bb} Time-resolved comparison of bond breakage (a-c) and bond formation (d-f) as a function of the measured stress $\sigma$ under LAOS deformation for the different 1 component gel concentrations: a,d) $\phi=7.5\%$, b,e) $\phi=10\%$, and c,f) $\phi=15\%$. The different colors refer to the key amplitudes referred to in the main text. The black arrows denote direction of curve throughout the oscillation.}
\end{figure*} 

In our simulations, bonds between neighboring gel particles can dynamically break and reform over time as the system is sheared. An overview of the statistics of these breakages and formations can be seen in Fig.~\ref{fig:bb} as a function of the shear stress induced in the samples by the imposed deformation. In each panel of the figure, the number of bonds broken or formed at a given point in time is represented as a percentage of the total bonds in the previous timestep.

Very little change in the number of bonds, either bond breakage or formation, can be seen at small amplitudes, and the changes are very small throughout the oscillation even at the peak in $G'$. At larger applied amplitudes, the percentages of bonds broken and/or formed in each time-step remain small, with the largest change being $\approx0.55\%$ per 1/100th of the period for all three gels, and noticeably less for smaller volume fractions. For the softest gels ($\phi=7.5\%$) the bond statistics data support the idea that the hysteresis indicated by the butterfly pattern in Fig.~\ref{fig:nemord} (far left plot) cannot be due to a significant restructuring of the bond network, and must be related, instead, to the non-affine motion underlying the re-alignment of the structure under shear, as seen in previous work \cite{colombo14,bouzid18}. Hence the emerging picture is that these very soft and sparsely connected structures allow for a hysteretic behavior of $S$ which is predominantly {\it elastic} in nature and due to large non-affine rearrangements, such as buckling of parts of the network, for example, that occur in the non-linear regime and are reminiscent of similar phenomena in liquid crystal elastomers \cite{warner}. 

Across all 1-component gels, the bond breakage and formation appear to be correlated to the shear stress. All curves are roughly centered on the point where $\sigma=0$ and seem to become more symmetric with increasing the gel volume fraction. The fact that a curve is not perfectly symmetric here suggests the presence of stresses built-in in the initial structure, and the data with increasing $\phi$ overall indicate that the more heterogeneous structures of the less dense gels may be more likely to have stress heterogeneities, whereas better symmetry is achieved in the gels at larger volume fractions, where the initial structure is more uniform (Fig.~\ref{fig:struct}). 

The bond statistics curves feature a pronounced hysteretic behaviour beyond the peak in $G'$, indicating that more bonds break or form when the magnitude of the stress is increasing as opposed to decreasing during yielding and as the gels start to flow. This hysteretic behavior is likely at the origin of the butterfly pattern found for the overall alignment $S$ of the gel structures at the same large amplitudes (see Fig. \ref{fig:nemord}), i.e. past the peak in $G'$. The findings confirm the idea that there are two distinct microscopic origin for the hysteresis in $S$ and for the buttlerfly patterns in Fig. \ref{fig:nemord}: one is rather determined by the presence of significant non-affine motion in very soft gel structures that are predominantly nonlinear elastic, and the other is associated instead to the rate at which the bond network can be restructured through plastic processes (i.e. bond breaking and formation) which limits the recovery of the overall alignment and of the rigid connected structure when reversing the strain (see also Figs.~\ref{fig:laos}, \ref{fig:spp}, and ~\ref{fig:nemord}).  

For the broken bonds, the pronounced hysteretic behavior in Fig.~\ref{fig:bb} indicate that fewer bonds may be "breakable" in the stages where stress is decreasing, suggesting that not all bonds in the gel networks are equally likely to break. The increase of the bond formation with the shear stress is more gradual than the bond breakage and actually starts prior to reaching $\sigma=0$, where the bond formation is non-zero in all cases. As the shear stress increases, however, the bond formation starts lagging slightly behind the bond breakage, which continues as the stress peaks and decreases. The maximum percentage of bond formation is also lower than the maximum percentage of bond breakage. While we would expect this to happen during yielding and flow, it also indicates that the conditions leading to bond formation are not exclusively determined by the stress, but are additionally related to the strain and/or rate \cite{colombo14} (hence the non-zero formation at $\sigma=0$). These findings, in our view, also point to differences in the availability of bonding sites (in terms of their spatial distributions), which changes with strain and rate, producing a lag near the maximum stress. The trends between amplitudes established for the bond breakage (i.e. increasing symmetry, maximum statistics, and hysteresis) appear to hold for bond formation as well.

Finally we note that the larger volume fraction gels, 
having a more pronounced yield stress fluid character according to our rheological analysis, spend a much larger portion of the period in the high stress region, which should increase the number of bonds broken or formed in this region. This is confirmed by the increased maximum percentages and the increase in hysteresis seen here.

\subsubsection{\label{sec:dsnad} Dynamic Structure and Non-affine Displacement}

\begin{figure*}
\includegraphics{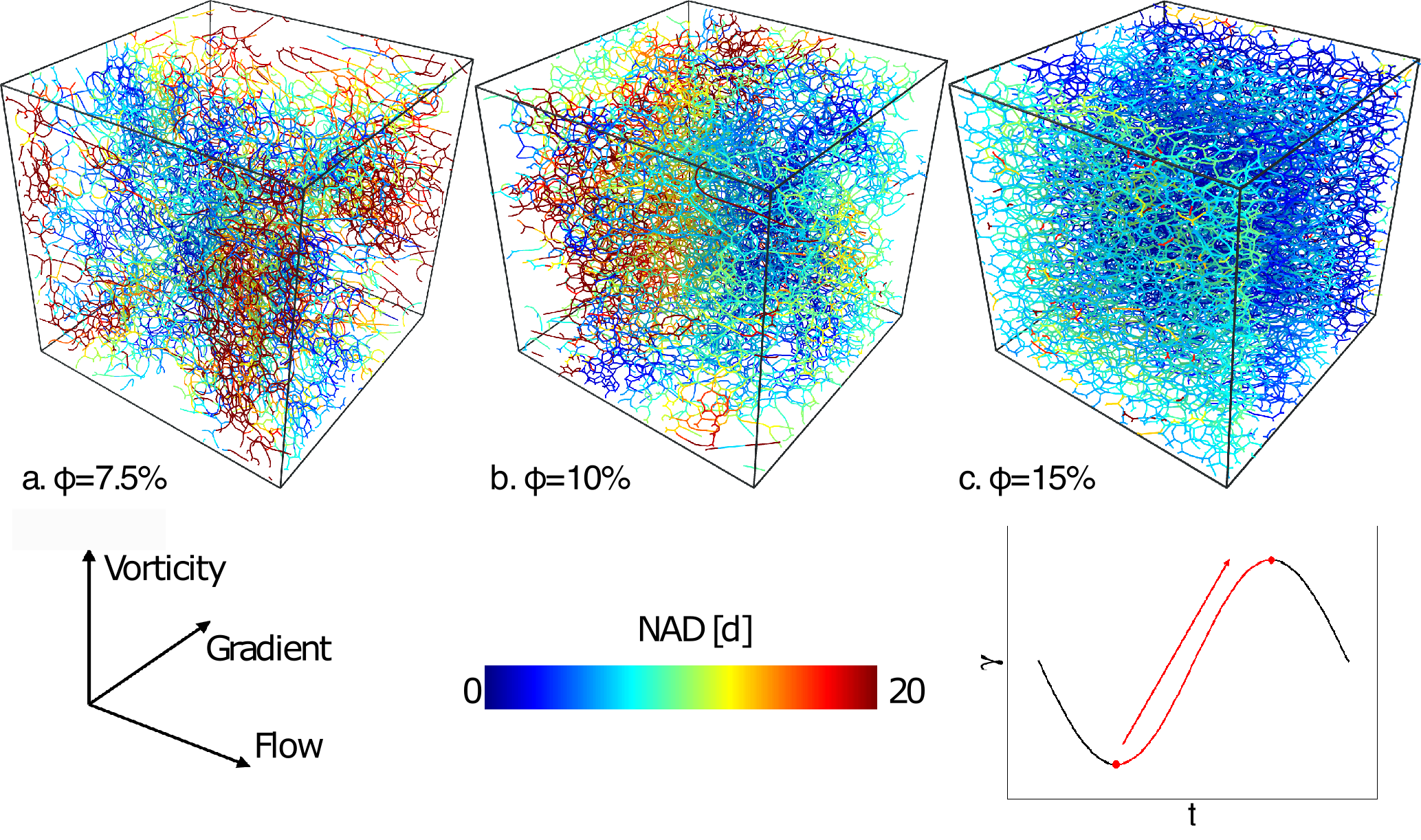}
\caption{\label{fig:nadmap} Maps of the accumulated non-affine displacement between strain extrema (i.e. from $\gamma=-\gamma_0$ to $\gamma=\gamma_0$) at the crossover amplitude (i.e. $G' = G''$) for the different 1 component gel concentrations: a) $\phi=7.5\%$, b) $\phi=10\%$, and c) $\phi=15\%$. }
\end{figure*}

In Fig.\ref{fig:struct} we have shown examples of the static structures of the 1-component gels at rest. The changes through the strain cycle of the nematic order parameter ($S$), of the bond statistics, and of the  nonlinear stress, suggest that the dynamical evolution of the gels microstructure may be associated to significant non-affine rearrangements throughout the period. While the non-affine motion in the pre-yielding regime has been investigated in previous work \cite{colombo14,bouzid18}, here we focus on the crossover between $G'$ and $G''$, i.e. on the post-yielding regime, to see if the apparently different behaviors detected in the rheology and in the other microstructural quantities are accompanied by qualitative differences in the non-affine motion. In order to understand the structural changes brought about via deformation, we focus on the strain extrema of the oscillation ($\gamma(t)=\gamma_0$), since the shear rate ($\dot{\gamma}$) is instantaneously zero, and the material is at its most structured point in the oscillation as a result. In the following we therefore compare the strain extrema the moduli crossover amplitude ($\gamma_0=\gamma_{crossover}$, orange curves in Figs.\ref{fig:laos}-\ref{fig:bb}) as the system goes through larger microstructural changes after it yields. Figure \ref{fig:nadmap} shows the structure at this point for each of the 1-component gel formulations, with the colormap denoting the non-affine displacement amplitude which will be discussed later in this section.  

In the $\phi$ = 7.5$\%$ case, there are several regions of the structure where large voids have opened in the sample, which explains this sample’s enhanced viscous behavior, and a number of clusters have densified. The $\phi$ = 10$\%$ structure does have a few smaller voids in the structure with less-well defined clusters, while the $\phi$ = 15$\%$ is relatively uniform microstructurally. In all cases, the application of shear seems to enhance any non-uniformities which existed in the structure at rest, given that the lower volume fraction gels are more heterogeneous to begin with.

We map the accumulated microscopic non-affine displacement (NAD) between strain extrema $-\gamma_0$ and $\gamma_0$ on top of the structure in Fig.\ref{fig:nadmap}. The non-affine motion, here quantified by the NAD, is one of the main sources of localization of flow and yielding in gels under deformation \cite{colombo14}, and, by comparing the NAD between the strain extrema, we can examine the cumulative effect of the shear through one iteration of the sequence of processes discussed above (section \ref{sec:spp_results}). As the concentration of particles increases, the maximum NAD decreases and the regions of high NAD become more clustered. In all cases, regions of nearly zero NAD are also present, but they evolve from being sparse clusters at $\phi$ = 7.5$\%$ to stacks along the vorticity direction at $\phi$ = 10$\%$, to hardly visible layers perpendicular to the gradient direction at $\phi$ = 15$\%$. This transition from homogeneously distributed and sparse clusters to anisotropic extended structures as layers seems to correlate with the changes in the character of the rheology as the volume fraction increases. For the softer gels, the heterogeneity in the NAD suggests some form of mixing accompanying the flow, while the layers in the $\phi$ = 15$\%$ gel suggest some form of shear-banding is present at larger volume fractions. 

\subsection{\label{sec:1compsum} Summary of 1-Component Gel Results}

The structure, dynamics, and rheological response of the 1-component gels under shear vary substantially with the volume fraction of the sample across all amplitudes of deformation. The time-resolved rheological analysis developed here shows how, in the linear regime, the ratio of elasticity to viscosity increasese as the gel density increases and the structure moves from weakly-connected to a more uniform mesh (Fig. \ref{fig:struct}a,b). This can be seen in the linear moduli (Fig.~\ref{fig:asweep}) and the relative moduli values for the "linear" amplitude (Fig.~\ref{fig:spp}).

In the intermediate amplitudes where non-linear elasticity dominates, the overall alignment (Fig.~\ref{fig:nemord}) and bond statistics (Fig.~\ref{fig:bb}) support the idea, suggested by the time-resolved rheological analysis, that the increase in the gel density and connectivity reduces the extent of the non-linear elasticity achievable below yielding. This result in a reduction of the $G'$ overshoot in Fig.~\ref{fig:asweep} and a distortion in the "peak $G'$" Lissajous curve (Fig.~\ref{fig:laos}).

At large amplitudes where nonlinear effects dominate, the increase in volume fraction leads to markedly different rheological behaviors, ranging from a quasi-linear viscoelastic fluid at low volume fractions to a yield stress fluid at higher concentrations, as seen in the relative shapes of the Lissajous curves (Fig.\ref{fig:laos}) and time-dependent Cole-Cole plots (Fig.\ref{fig:spp}). When inspecting for the structural cause of this shift, it is apparent that the distribution of particles and bonds changes quite substantially at large deformations (Fig.\ref{fig:bb}), leading to drastically different types of non-affine motion (Fig.~\ref{fig:nadmap}). At smaller volume fractions, the sample appears to break into tightly bound clusters with only transient connections between them, akin to a fluid composed of elastic gel blocks in a viscous medium. At larger $\phi$, the structure remains more spatially uniform (akin to the unsheared gels in Fig.\ref{fig:struct}), however larger and anisotropic domains where non-affine motion concentrate emerge. All these changes seem to be associated to the transition in the gel architecture from sparse to more tightly and uniformly connected \cite{colombo14,bouzid18}. 

\section{\label{sec:2comp} Results for 2-Component Gels}

In order to understand the impact of adding the second component into the gels, we will compare the rheology and structure of the 2-component gel formulations to the $\phi$ = 10$\%$ and $\phi$ = 15$\%$ 1-component gels, as they have similar overall total volume fractions. In the 2-component systems, we denote $\phi_1$ as the volume fraction of primary component (component 1, corresponding to the 1-component gels studied previously) and $\phi_2$ as the volume fraction of secondary component (component 2, which does not form a gel network on its own, as described in section \ref{Methods}). All component ratios will be in the form $\phi_1:\phi_2$.

\subsection{\label{2c_rheo} Comparison of Rheology}

\begin{figure}
\includegraphics{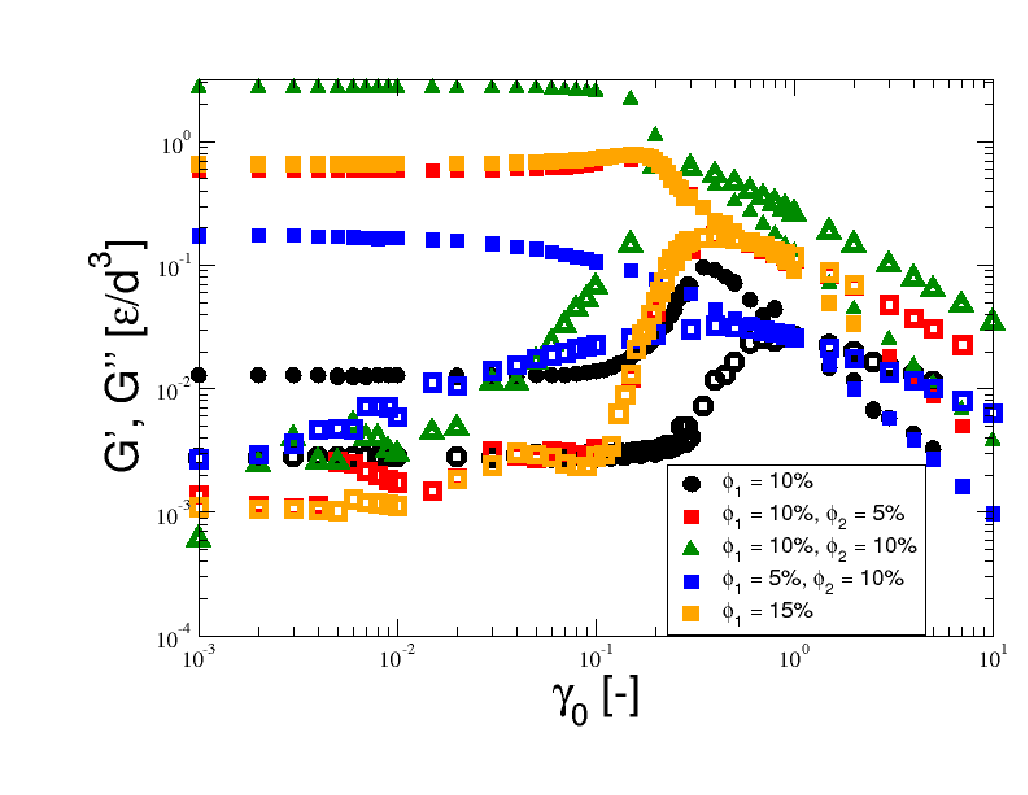}
\caption{\label{fig:asweep_2c} Comparison of the amplitude sweeps across the 1-component and 2-component gels. The different colors denote the different gel formulations, while the different shapes denote the total volume fraction of both components.}
\end{figure}

\begin{figure}
\includegraphics{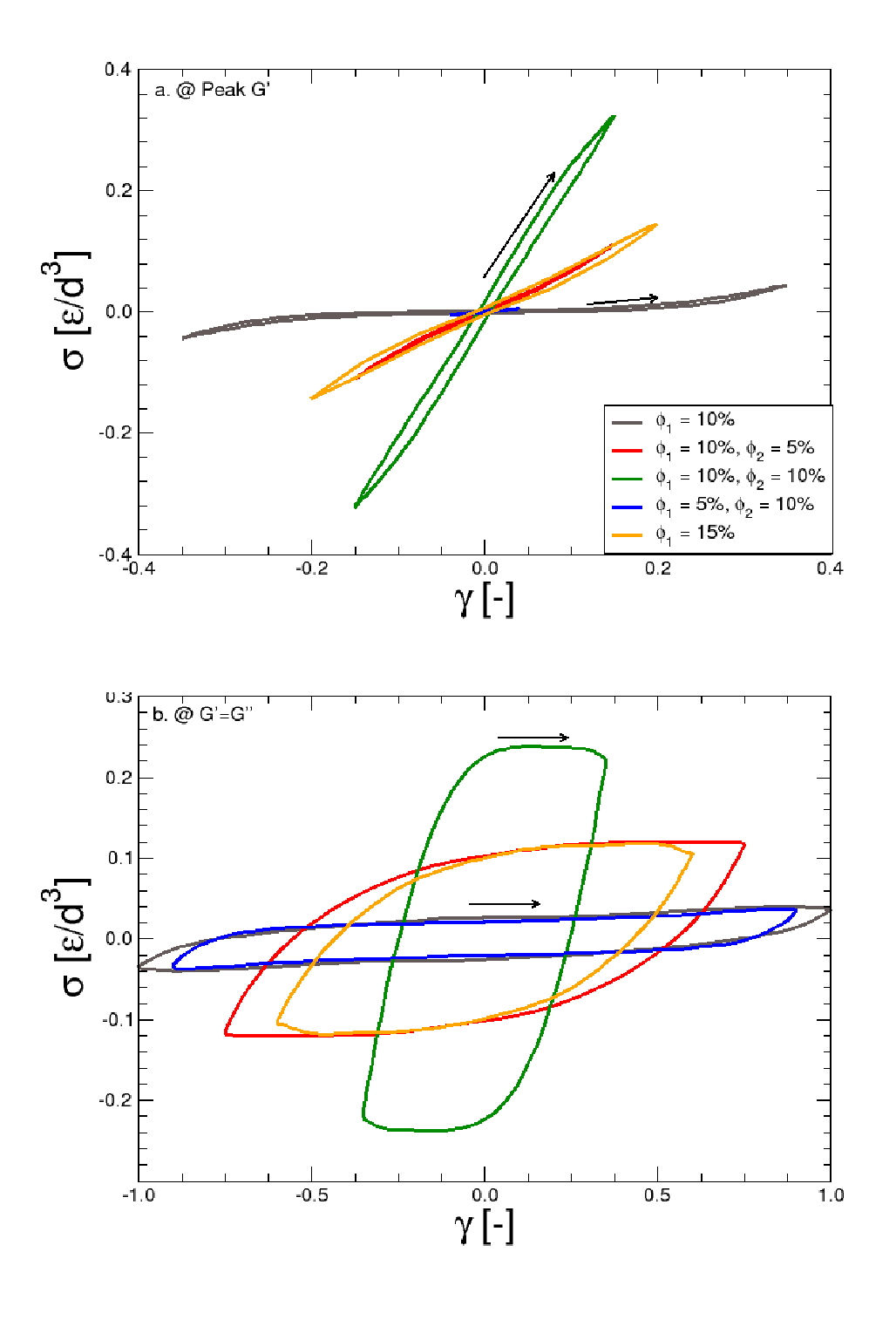}
\caption{\label{fig:laos_2c} Comparison of the elastic Lissajous curves of the time-resolved LAOS across the 1-component and 2-component gels at: a) the peak in $G'$ (or peak equivalent) and b) the crossover point (i.e. $G' = G''$). The different colors denote the different gel formulations. The black arrows denote direction of curve throughout the oscillation.}
\end{figure}

\begin{figure}
\includegraphics{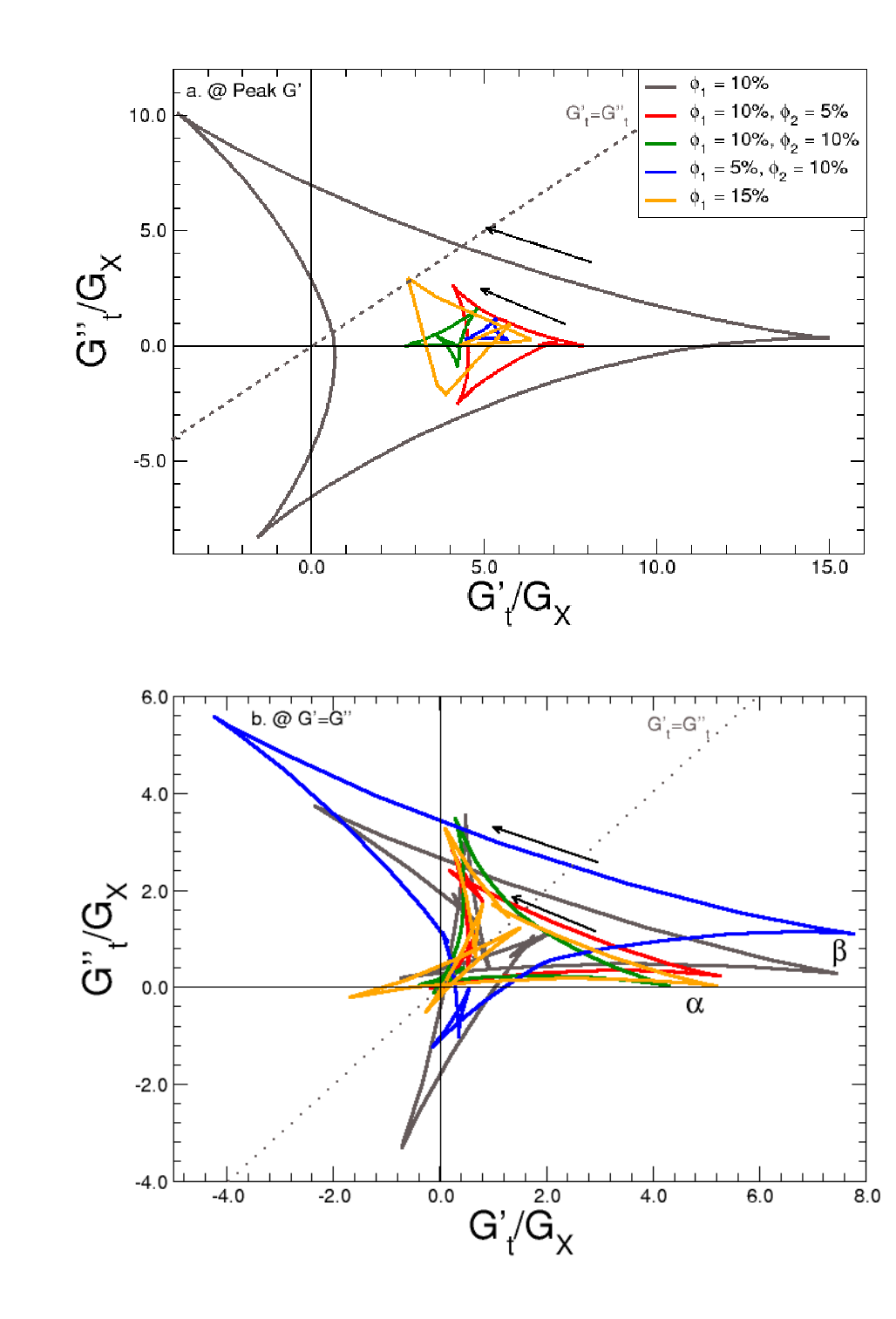}
\caption{\label{fig:normspp_2c} Cole-Cole plots of the time-resolved dynamic SPP moduli ($G^\prime_t(\omega,t)$ and $G^{\prime \prime}_t(\omega,t)$) across the 1-component and 2-component gels at: a) the peak in $G'$ (or peak equivalent) and b) the crossover point (i.e. $G' = G''$). The moduli are normalized by the crossover modulus ($G_X = G'_{crossover} = G''_{crossover}$) The different colors denote the different gel formulations. The black arrows denote direction of curve throughout the oscillation. The different groups of trajectories in b) are denoted by $\alpha$ and $\beta$, as described in the text.}
\end{figure}

A comparison of the amplitude sweep behavior of the selected gel formulations can be seen in Fig.~\ref{fig:asweep_2c}. It is immediately apparent that the $\phi_1=10\%, \phi_2=5\%$ formulation (red, 2:1 $\phi_1:\phi_2$) shows nearly the same behavior across the amplitude sweep as the 1-component gels at $\phi$ = 15$\%$, with overshoots in both moduli and a large amplitude crossover. The moduli in the 2:1 formulation are slightly smaller at most amplitudes, probably due to the similar total particle volume fraction, in spite of the presence of the second component.

The $\phi_1=10\%, \phi_2=10\%$ formulation (green triangles, 1:1 $\phi_1:\phi_2$) shows larger moduli than the other formulations across the majority of the sweep, due to the larger overall volume fraction ($20\%$). Noticeably, it does not show any overshoot in $G'$, though an overshoot is still seen in $G''$. Rheologically, this behavior can be classified as weak strain overshoot amplitude sweep behavior \cite{hyun11}, which is akin to what is seen in simple yield stress fluids \cite{donley20}.

Finally, for the $\phi_1=5\%, \phi_2=10\%$ formulation (blue squares, 1:2 $\phi_1:\phi_2$), the moduli across the sweep are lower than the other 2-component formulations. The transition from the linear regime also occurs much earlier and more gradually than in the other formulations. While the 1:2 gel shows no $G'$ overshoot, above the moduli crossover its response appears to follow quite closely the one of the $\phi$ = 10$\%$, 1-component gel. Given the larger total volume fraction, this behavior support the idea that the contribution of the secondary component to the non-linear properties of the material is weaker than the primary component, especially as it becomes a larger fraction of the total sample. This similarity in the rheological response of the 1:2 formulation and the 1-component gel at a different volume fraction appears only above yielding, highlighting how the same post-yielding behavior can, in principle, be attained in spite of starting from quite different microstructures at rest.

The critical amplitudes for the 2-component gels, as discussed for the 1-component gels in section \ref{sec:1comp}, are listed in Table \ref{tab:2comp}. In the following, we'll compare the time-resolved rheology of the 1-component and 2-component gels at the peak in $G'$ and at the crossover point, i.e. well into the non-linear behavior of the materials. In the case of the 1:1 and 1:2 2-component formulations, which do not have a peak in $G'$, we will use the amplitude at which the linear and non-linear regime trends in $G'$ intersect.

\begin{table}
\caption{\label{tab:2comp} Applied strain amplitudes used for rheology and structure comparisons between the 2-component gel formulations. The same definitions are used as in Table \ref{tab:1comp} except where noted. All amplitudes are given in strain units.}
\begin{ruledtabular}
\begin{tabular}{llll}
 &$\phi_1$:$\phi_2$=2:1&$\phi_1$:$\phi_2$=1:1&$\phi_1$:$\phi_2$=1:2\\
\hline
linear & 0.001 & 0.001 & 0.001\\
non-linear & 0.03 & 0.02 & 0.005\\
peak $G'$ & 0.15 & 0.15\footnotemark[1] & 0.04\footnotemark[1]\\
$G'$=$G''$ & 0.75 & 0.35 & 0.90\\
$G''$>$G'$ & 1.0 & 0.45 & 1.2\\
\end{tabular}
\end{ruledtabular}
\footnotemark[1]{These sweeps do not have a peak in $G'$. As such the $G'$ trend intersection as described in the text was used instead.}
\end{table}

The elastic Lissajous curves of the LAOS at these two amplitudes across the different formulations can be seen in Fig.~\ref{fig:laos_2c}. At the peak in $G'$ or peak equivalent (Fig.~\ref{fig:laos_2c}a), nearly all the amplitudes show predominantly elastic behavior. The $\phi$ = 10$\%$ 1-component formulation is the only one to show strong non-linear elasticity at this point, though small amounts of non-linear elasticity can be seen in the $\phi$ = 15$\%$ 1-component and 2:1 2-component formulations, supporting the idea that the addition of the second component tend to reduce the non-linear elasticity in the composite gels.

At the crossover point (Fig.~\ref{fig:laos_2c}b), all formulations have a similar parallelogram-like shape, however the aspect ratio does become more vertical as the value of the crossover modulus (i.e. $G_X = G' = G''$) increases. Notably, the $\phi$ = $15\%$ and 2:1 gels show nearly the same time-resolved LAOS behavior, as do the $\phi$ = $10\%$ and 1:2 formulations, hence their very similar average moduli in this portion of the amplitude sweep (Fig.~\ref{fig:asweep_2c}) are the result of their similar time-resolved behavior. Overall this comparison indicate that the relative composition of the mixtures changes dramatically the nonlinear response of the gels, similar to what already observed for the linear response \cite{vereroudakis20}.

For the time-resolved rheological analysis, given the difference in the values of the moduli across the different gels, we normalize the time-resolved SPP moduli (Fig.~\ref{fig:normspp_2c}) by their crossover modulus $G_X$ to allow for a more direct comparison. At the peak in $G'$ or peak equivalent (Fig.~\ref{fig:normspp_2c}a), $G'_t > G''_t$ for all formulations except for the $\phi$ = $10\%$, which also shows by far the strongest relative variation in SPP moduli at this point. We note that the two composite formulations without a peak in $G'$ (i.e. the 1:1 and 1:2) show the least relative variation in the time resolved moduli at their peak equivalent amplitude. At the crossover point (Fig.~\ref{fig:normspp_2c}b), the normalized time-dependent moduli fall roughly along one of two trajectories: 1) a more reduced trajectory (denoted by $\alpha$ in Fig. \ref{fig:normspp_2c}) traced by the $\phi$ = $15\%$, 2:1, and 1:1 gels; and 2) a much more extended trajectory (denoted by $\beta$ in Fig. \ref{fig:normspp_2c}) traced by the $\phi$ = $10\%$ and 1:2 gels. The first of these trajectories indicates that minimal elastic recoil and backflow are occurring within the period, since it is nearly completely in the first quadrant, while the second has significant regions of both. The smaller relative variations in the moduli through the first trajectory also suggests that the yielding behavior overwhelms any non-linear elasticity present in the gels, while the formulations which follow the second trajectory may still experience some non-linear elasticity in conjunction with yielding.

\subsection{\label{2c_struct} Comparison of Structural Measures}

\begin{figure}
\includegraphics{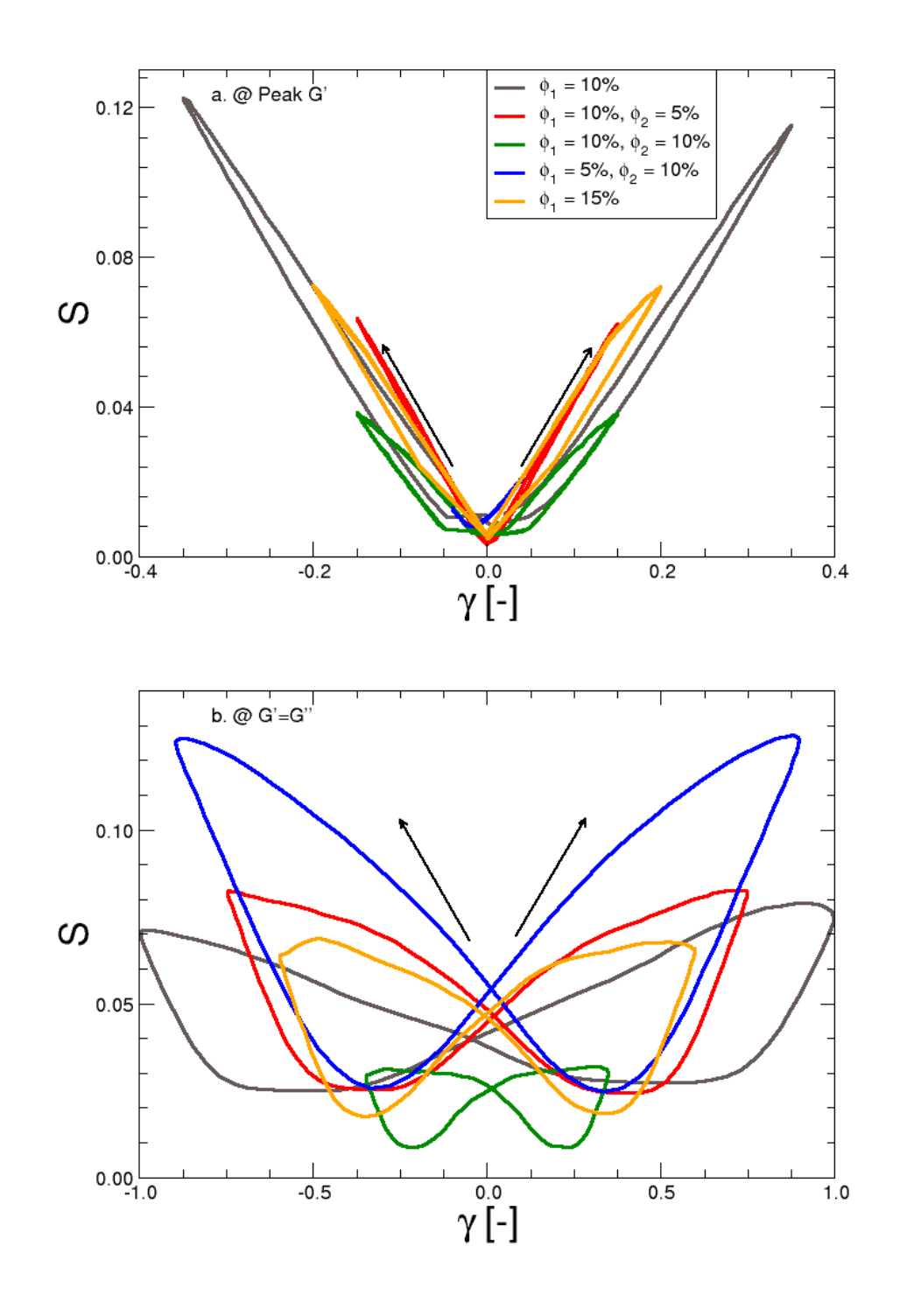}
\caption{\label{fig:nemord_2c} Comparison of the nematic order parameter $S$ with the strain $\gamma$ under LAOS deformation across the 1-component and 2-component gels at: a) the peak in $G'$ (or peak equivalent) and b) the crossover point (i.e. $G' = G''$). The different colors denote the different gel formulations. The black arrows denote direction of curve throughout the oscillation.}
\end{figure}

\begin{figure}
\includegraphics{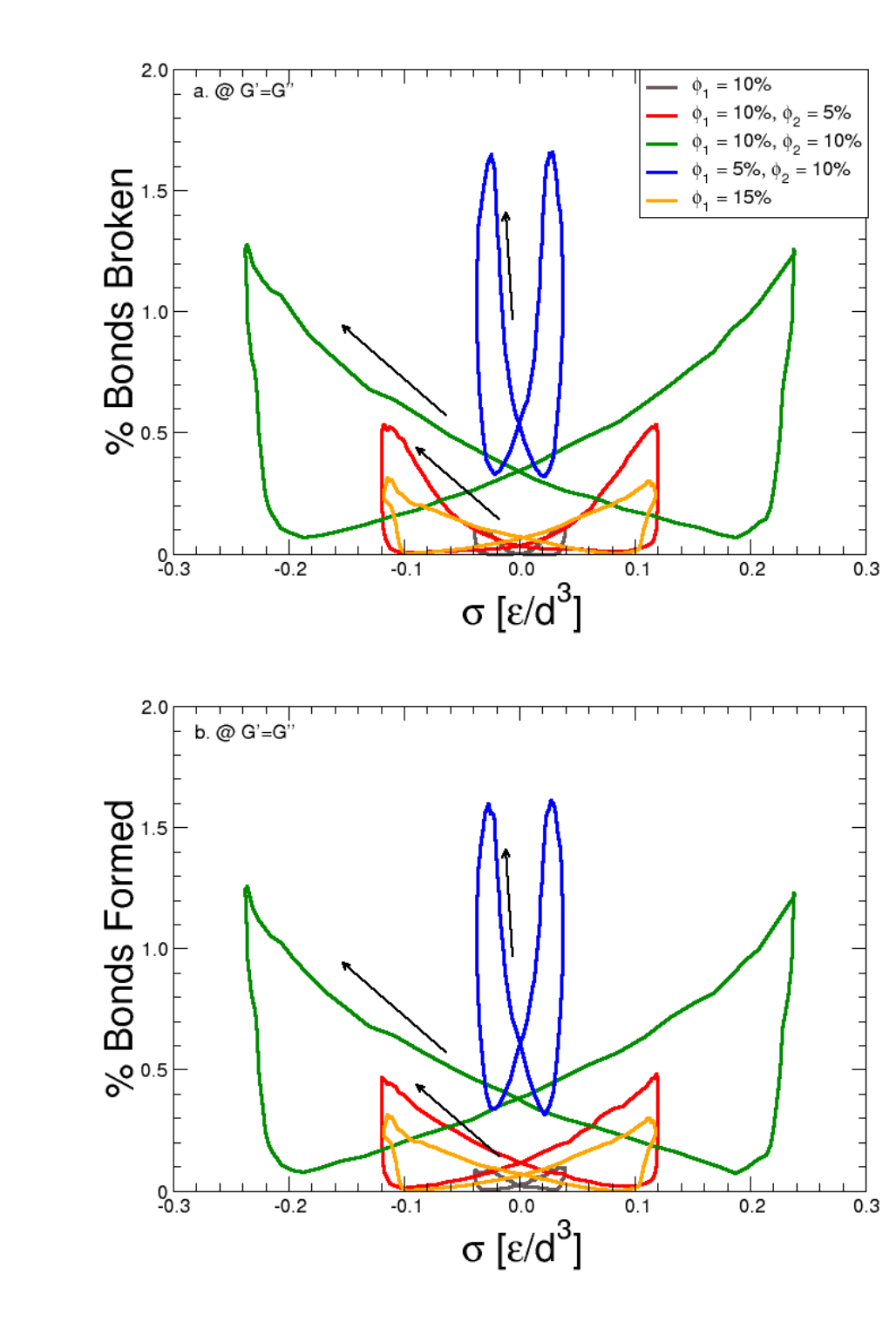}
\caption{\label{fig:bb_2c} Time-resolved comparison of bond breakage (a) and bond formation (b) as a function of the measured stress across the 1-component and 2-component gels at the crossover point (i.e. $G' = G''$). The different colors denote the different gel formulations. The black arrows denote direction of curve throughout the oscillation.}
\end{figure}

\begin{figure*}
\includegraphics{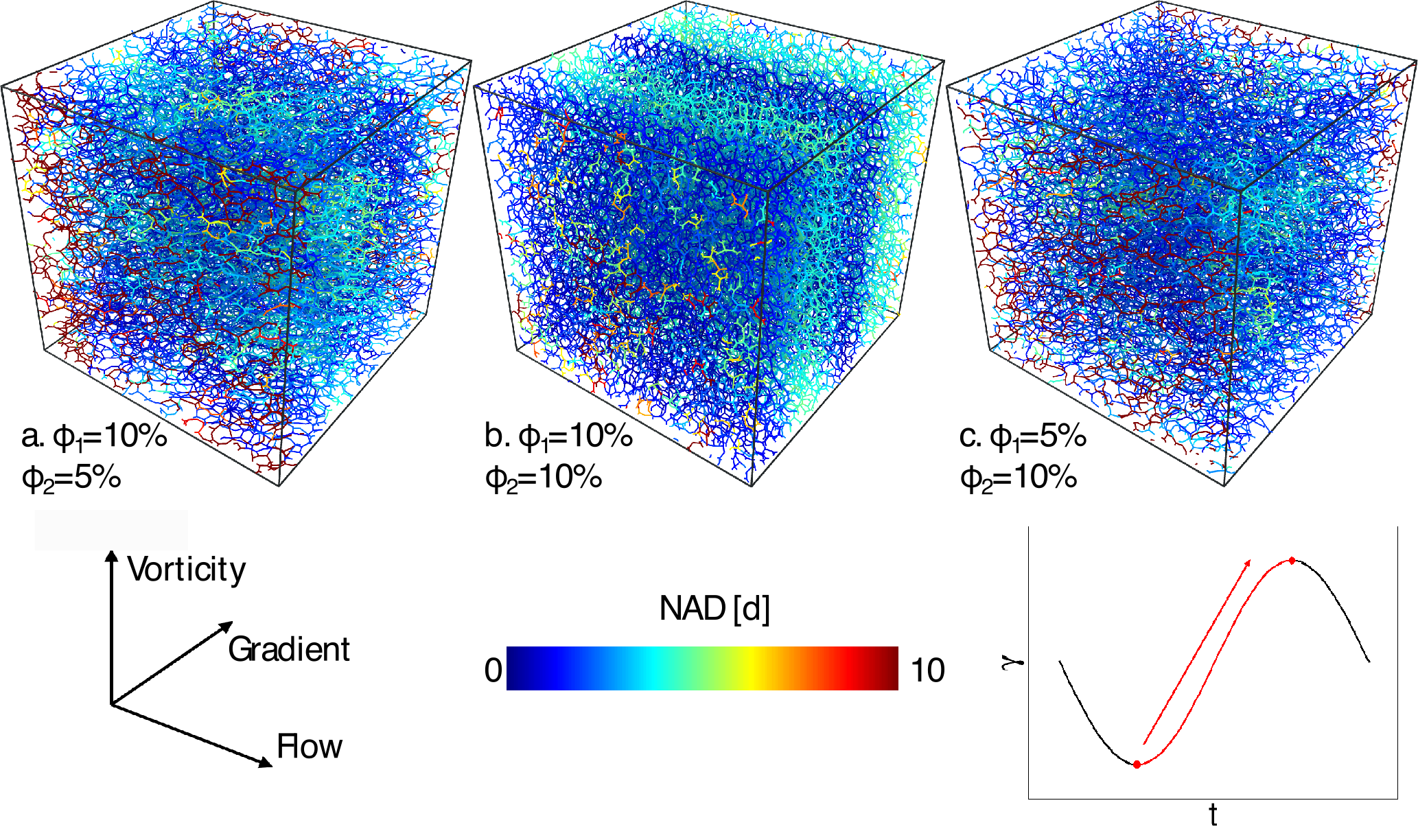}
\caption{\label{fig:nadmap_2c} Maps of the accumulated non-affine displacement between strain extrema (i.e. from $\gamma=-\gamma_0$ to $\gamma=\gamma_0$) at the crossover amplitude (i.e. $G' = G''$) for the different 2-component gel formulations: a) $\phi_1=10\% + \phi_2=5\%$, b) $\phi_1=10\% + \phi_2=10\%$, and c) $\phi_1=5\% + \phi_2=10\%$.}
\end{figure*}

Moving to rheo-structural comparisons of the 1- and 2-component gels, we first consider the shear-induced alignment (Fig.~\ref{fig:nemord_2c}) as discussed in section \ref{sec:shinani}, again focusing on the peak in $G'$ (or peak equivalent) and the crossover amplitude. Looking at the peak amplitude (Fig.~\ref{fig:nemord_2c}a) we see the same v-shaped pattern, with a negligible hysteresis, seen in the 1-component gels before yielding, directly linking the microstructural alignment to the strain imposed at these amplitudes. The slopes are comparable in spite of the composition and microstructural differences but the degree of alignment attained and the point at which the materials yield or breaks down vary across the different gels. Overall the addition of second component seem to always decrease the capacity of the gels structure to develop a pronounced degree of alignment, however this reduction seems similar across the different mixtures.

At the crossover amplitude (Fig.~\ref{fig:nemord_2c}b), $S$ shows butterfly shapes similar to those observed in the 1-component gels in the same conditions (Fig.~\ref{fig:bb}). The range in $S$ appears to scale inversely with the crossover modulus for all gels, i.e. gels that remain stiffer upon yielding also tend to have a reduced variation range of $S$ over the cycle, except for the $\phi$ = $10\%$ 1-component system, where instead $S$ varies less compared to the 2-component gels with total higher volume fraction. Hence it seems that the addition of component 2, no matter in which percentage, always makes the gel structure more able to align under shear. We note however that, while the $\phi$ = $15\%$ and 2:1 gels appear to have nearly the same trajectory in Fig.~\ref{fig:nemord_2c} (bottom plot), the 1:2 2-component gel appears to show significantly more alignment than the $\phi$ = $10\%$ 1-component system, in spite of the similarity between those two systems in both the time-averaged (Fig.\ref{fig:asweep_2c}) and time-resolved rheology (Fig.\ref{fig:laos_2c}b $\&$ Fig.\ref{fig:normspp_2c}b). These observations, therefore, suggest that the relative composition of the mixtures in the composite gels, i.e. the relative amount of component 2 introduced over the same total volume fraction, can still lead to different microstructures upon flow in a more subtle but potentially significant way, that is not necessarily evident in the macroscopic rheological signal.

 When comparing bond breakage and formation (Fig.\ref{fig:bb_2c}) at the crossover point of the amplitude sweeps ($G'=G''$), the 2-component gels display larger changes in the percentage of bonds broken or formed each 1/100th of the period, which may be due to the fact that overall the aggregates in the second component are weaker by construction, due to weaker microscopic interactions in the model (see Section ~\ref{Model}). {We note that} for the  2:1 gels, the statistics are only slightly larger than the $\phi$ = $15\%$ 1-component gel, however, as the relative amount of component 2 is increased, the percentages of bonds both broken and formed go  up substantially. The bond breaking and formation data clearly demonstrate that the 2:1 composite gel features a much higher amount of broken and formed bonds with respect to the $\phi=10\%$ 1-component gel, suggesting that this is the sources of the much higher degree of alignment shown in Fig.~\ref{fig:nemord_2c}. This insight clarifies that the similarity of the shapes of Lissajous curves (Fig.~\ref{fig:laos_2c}) may still correspond to a pronounced structural difference, which may be instead revealed by the time-dependent rheological analysis showing a reduction of the trajectory in the Cole-Cole plot upon yielding for the $\phi=10\%$ 1-component  gel and an expansion, instead, for the 2:1 composite one (Fig.~\ref{fig:normspp_2c}). Hence such opposite trend could signal the markedly more dynamical nature of the structures in the 2:1 composite gel. Finally, we note that bond breaking and formation occurs at significantly different stresses when changing the mixture compositions, therefore demonstrating that the architecture of the gel network in the component 1, which is significantly modified by changing the relative amount of component 2 for the same total volume fraction \cite{vereroudakis20}, also changes the reconfigurability of the gels.

The gel structures under shear and the accumulated non-affine displacement in the 2-component gels are shown in Fig.\ref{fig:nadmap_2c}, using the same conditions as defined in section \ref{sec:dsnad}. The structures are all relatively uniform, consistent with the fact that their total volume fractions are all the same as, or larger than, the $\phi$ = $15\%$ 1-component gel which showed a similar, uniform structure for the same shear conditions. Interestingly, all of the 2-component gels show significantly smaller values of the accumulated NAD (note that the colorbar in Fig.\ref{fig:nadmap_2c} runs from 0-10$d$, as opposed to 1-20$d$ in Fig.\ref{fig:nadmap}),  indicating that the more dynamical nature of the second component, together with decreasing the non-linear elasticity and providing more capability to align under shear, also favors more affine behavior upon entering the flow regime. The 1:2 gel (Fig.\ref{fig:nadmap_2c}c) appears to feature the least nonaffine motion, with only small portions showing any measurable NAD, indicating that both the larger overall alignment (Fig.\ref{fig:nemord_2c}b) and bond breakage/formation (Fig.\ref{fig:bb_2c}b) just discussed may be at the origin of this. Finally, we also note that the 1:1 gel (Fig.\ref{fig:nadmap_2c}b) shows some layering perpendicular to the gradient direction, similar to that seen in the $\phi$ = $15\%$ 1-component gel, and more pronounced than in the 2:1 system (Fig.\ref{fig:nadmap_2c}a), pointing to some differences possibly controlled by the different architecture of the network in component 1 in the two mixtures.

\subsection{\label{2c_summary} Summary of 2-Component Gel Results}

The rheology of the 2-component gels, and the evolution of their structure upon yielding, appears to be the result of two main causes. The first is the more dynamical nature of the second component, which reduces non-linear elasticity, increases the capacity to align under shear, which is associated to a much weaker presence of non-affine motion. The majority of the behavior seen for both the 2:1 ($\phi_{1+2}$ = 15$\%$) and 1:1 ($\phi_{1+2}$ = 20$\%$) gels, as they act as a natural, but more dynamical, continuation of the pattern established for the 1-component gels above (section \ref{sec:1compsum}). The second is the fact that the architecture of the network in component 1 can be significantly modified by changing the relative amount of component 2 for the same total volume fraction, and this leads to a number of changes observed when comparing the 1:2 ($\phi_{1+2}$ = 15$\%$) gel to the 2:1. As the fraction of secondary component increases, the stresses at which bonds break and form, not only their total amount, change significantly, resulting in qualitatively different rheological responses due to different amount of alignment and characteristics in the nonaffine motion.  

\section{\label{sec:conclu} Conclusions and Outlook}

We have performed numerical LAOS tests on model particulate gels, using coarse-grained nonequilibrium molecular dynamics simulations. The gels investigated span a range of different linear and non-linear behaviors when we change the particle content and introduce a second component that can be interspersed in the first network. We have complemented the rheological analysis developed in \cite{rogers11,rogers17} specifically for LAOS tests with an analysis that relies on microscopic observables quantifying the statistics of bonds formed/broken during the strain cycles, the overall bond alignment, and the non-affine motion. Through this combination, we demonstrate the versatility of the SPP approach and use the microscopic information to confirm or deepen the insight provided by the SPP on the changes in the material properties during the strain cycles. We show the time-resolved SPP analysis is sensitive to changes of the gel network architectures, well beyond the averaged moduli amplitude, and, through the coarse-grained molecular dynamics simulations, we identify the microscopic sources of nonlinear elastic and plastic behaviors, clarifying in which deformation regimes they can lead to specific consequences for the rheological signals.  

When analyzing the gels made of one component, we have found that relatively small changes of the modulus in the linear regime, due to a change in the particle volume fraction, may correspond to dramatic differences in the way the materials yield. In particular, for gels that are initially weaker and more sparsely connected, large deformations can be accumulated with negligible bond breaking but significant nonaffine motion before yielding, which also persist in sparse regions as they start to flow, and feature pronounced nonlinear elasticity and capacity to align under shear. Gels that are denser, and more homogeneous to start with, tend instead to significantly reconfigure their structural networks, via bond breaking and formation, to yield, with more pronounced plastic behavior as they start to flow. The microscopic plastic processes are also the main source, in these cases, of pronounced alignment of the structures to the shear flow and of nonaffine motion. As a consequence, the nonaffine motion spacial maps seem less affected by the structural heterogeneities in the gel architecture, which appears predominant in the softer and less dense gels,  and are rather reminiscent of flow instabilities, such as shear banding, which may be mainly controlled by the geometry of the flow. These observations and the analysis performed here provide therefore potentially useful hints to design the nonlinear rheological response of gel materials.

A broader range of rheological properties can be obtained by varying the gel composition by adding a second component. The rheological response of these composite gels is determined of course by the physical chemistry of the two components, however recent experiments have demonstrated that the relative compositions of the mixtures can drastically modify those effects, due to the role played by the different gel architectures \cite{vereroudakis20}. Our combination of the time-resolved SPP analysis with the microscopic information provided by the simulations highlights that the introduction of a second, more dynamical, component, interspersed in the first network, dramatically change the rheological response primarily by reducing the nonlinear elasticity and promoting more reconfigurable and homogeneous micristructures upon yielding, where the nonaffine motion is significantly reduced. However, the effect of the second component may be modified by the architecture of the first gel networks, as we find that denser gels of the first component lead to composite gels that remain more elastic and are much less able to align under flow conditions. On the contrary, when the first network is more sparse and dilute, the second component leads to much more dynamical and reconfigurable structures that can follow more effectively the imposed deformation through relatively easier bond breaking and formation. We have produced maps of the magnitude of the accumulated non-affine displacements between the strain-extrema during the oscillations, which can give hints of when and where these materials may be prone to flow localization. The data suggest that the addition of the second component may promote flow localization along the gradient direction under shear, but also that this is a combined effect of the presence of the second component {\it and} the topology of the first network structure. Similar type of topological effects have been highlighted in experiments and simulations of the linear properties of such composite gels \cite{vereroudakis20} and our results here elucidate how such features persist and affect the non-linear regime, with consequences for yielding and flow.

The scope of the work discussed here is to demonstrate how combining microscopic simulations with LAOS tests and the time-resolved rheological analysis provided by the SPP can help gain new, deeper insight into the way particulate gel microstructures respond to large cyclic shear deformation, to predict shear induced restructuring, shear localization and failure.     
Future work will expand this approach to study memory, fatigue, and thixotropy.

\section*{Acknowledgements}

This work was funded by the NIST PREP Gaithersburg Program (70NANB18H151), ACS PRF and NSF DMREF (CBET-2118962). The authors thank Simon Rogers and Dimitris Vlassopoulos for insightful discussions.

\section*{References}
\bibliography{gjd_mb_edg_revision}

\end{document}